\documentclass[a4paper,fleqn,usenatbib]{mnras}

\usepackage[T1]{fontenc}
\usepackage{ae,aecompl}

\usepackage{graphicx}	
\usepackage{amsmath}	
\usepackage{amssymb}	

\DeclareMathOperator{\popcnt}{popcnt}

\newcommand{\mpcoh}{\,h^{-1}\,{\rm Mpc}}

\title[Unbiased clustering with DESI]{Unbiased clustering estimates with the DESI fibre assignment}

\author[Davide Bianchi et al.]{
Davide Bianchi,$^{1}$\thanks{E-mail: davide.bianchi@port.ac.uk},
Angela Burden$^{2}$,
Will J. Percival$^{3,4,1}$,
David Brooks$^{9}$,
\newauthor
Robert N. Cahn$^{5}$,
Jaime E. Forero-Romero$^{6}$,
Michael Levi$^{5}$,
Ashley J. Ross$^{7,8}$,
\newauthor
Gregory Tarle$^{10}$
\\
$^{1}$Institute of Cosmology \& Gravitation, Dennis Sciama Building, University of Portsmouth, Portsmouth, PO1 3FX, UK\\
$^{2}$Dept. of Physics, Yale University, New Haven, CT 06511, USA\\
$^{3}$Department of Physics and Astronomy, University of Waterloo, 200 University Ave W, Waterloo, ON N2L 3G1, Canada\\
$^{4}$Perimeter Institute for Theoretical Physics, 31 Caroline St. North, Waterloo, ON N2L 2Y5, Canada\\
$^{5}$Lawrence Berkeley National Laboratory, 1 Cyclotron Road, Berkeley, CA 94720, USA\\
$^{6}$Departamento de F\'isica, Universidad de los Andes, Cra. 1 No. 18A-10, Edificio Ip, Bogot\'{a}, Colombia\\
$^{7}$Center for Cosmology and AstroParticle Physics, The Ohio State University, 191 West Woodruff Avenue, Columbus, OH 43210, USA\\
$^{8}$Department of Astronomy, The Ohio State University, 4055 McPherson Laboratory, 140 W 18th Avenue, Columbus, OH 43210, USA\\
$^{9}$Department of Physics \& Astronomy, University College London, Gower Street, London, WC1E 6BT, UK\\
$^{10}$Physics Department, University of Michigan Ann Arbor, MI 48109, USA
}

\date{Accepted XXX. Received YYY; in original form ZZZ}

\pubyear{2017}

\begin{document}
\label{firstpage}
\pagerange{\pageref{firstpage}--\pageref{lastpage}}
\maketitle

\begin{abstract}
The Emission Line Galaxy survey made by the Dark Energy Spectroscopic Instrument (DESI) survey will be created from five passes of the instrument on the sky.
On each pass, the constrained mobility of the ends of the fibres in the DESI focal plane means that the angular-distribution of targets that can be observed is limited.
Thus, the clustering of samples constructed using a limited number of passes will be strongly affected by missing targets.
In two recent papers, we showed how the effect of missing galaxies can be corrected when calculating the correlation function using a weighting scheme for pairs.
Using mock galaxy catalogues we now show that this method provides an unbiased estimator of the true correlation function for the DESI survey after any number of passes.
We use multiple mocks to determine the expected errors given one to four passes, compared to an idealised survey observing an equivalent number of randomly selected targets.
On BAO scales, we find that the error is a factor 2 worse after one pass, but that after three or more passes, the errors are very similar.
Thus we find that the fibre assignment strategy enforced by the design of DESI will not affect the cosmological measurements to be made by the survey, and can be removed as a potential risk for this experiment.
\end{abstract}




\section{Introduction}

The observed clustering of galaxies provides a wide range of physical
measurements, including measuring the cosmological expansion rate
using Baryon Acoustic Oscillations (BAO) as a standard ruler, and
using Redshift-Space Distortions (RSD) to measure structure
growth. However, making such measurements relies on being able to
distinguish between fluctuations in galaxy density caused by intrinsic
fluctuations in the matter field, and those caused by experimental effects.
In particular, the design of the survey itself can lead to a distortion of the clustering signal
through the algorithm that selects which galaxies to observe from a complete parent sample of targets.

An example of experimental design affecting clustering measurements is
the decrement of angularly-close galaxy pairs in Sloan Digital Sky
Survey (SDSS) galaxy samples, caused by the physical limitation of not
being able to place two fibres close to each other within the focal
plane of the instrument \citep{masjedi2006, dawson2013}. 

For galaxy surveys made using the Dark Energy Spectroscopic Instrument
(DESI), there will be a similar issue, caused by limitations in the
placement of fibres close to each other within the focal plane\footnote{More precisely, the issue is the collision of the fibre positioners, which can happen even if the fibres are relatively far from each other.}
\citep{amir2016a, amir2016b}. In addition, each of the 5000 fibres is constrained to a ``patrol radius'', which is only a
very small fraction of the focal plane. Although each patrol region
overlaps that of adjacent fibres (about 15\% of the total area is within the patrol radius of two different fibres), the overall pattern means that
observations are more evenly spread than the parent sample. The scale
encoded from this selection is visible in the measured correlation
function as we explain in see Section~\ref{sec fa} (see also \citealt{burden2017, pinol2017}).
Due to these fibre-allocation issues, the maximum number of redshifts that can be measured from highly clustered groups of galaxies is set by the number of pointing overlaps.
In this respect, as DESI will cover its footprint with five passes of the instrument, the close pair effect in clustered groups will be less pernicious than for SDSS, despite the overall higher completeness of this latter survey.

In \citet{bianchi2017} we introduced a general algorithm to remove the
effect of only observing a limited subsample of galaxy targets, by
using the targeting algorithm itself to define a probability of
selection for each pair (a brief description of the algorithm is
provided in Section~\ref{sec model}). Depending on the way in which
targets are selected, the probability may have to include spatially
moving the survey region and/or altering random choices made within
the algorithm, as discussed in Section \ref{sec zero prob}.

In this paper we apply this algorithm to mock catalogues of the DESI
Emission Line Galaxy survey. Details of the mocks are given in
Section~\ref{sec mocks}. The resulting correlation function
measurements are presented in Section~\ref{sec results: xi}, and in
Section~\ref{sec results: err}, we compare the errors on the corrected
measurements to those from catalogues constructed from randomly
sampled targets, and against cosmic variance errors.
Note that as part of the algorithm, we include the angular pair up-weighting introduced
by \citet{percival2017}, which reduces the errors after applying the
weighting, by using our knowledge about the angular clustering of the
unobserved targets.

In Section~\ref{sec results: BAO} we investigate the effectiveness of the algorithm in recovering the position of the BAO peak. 
We end with a brief summary of our results in Section~\ref{sec conclusions}.

\section{DESI fibre assignment}\label{sec fa}

\begin{figure*}
 \begin{center}
   \includegraphics[width=16cm]{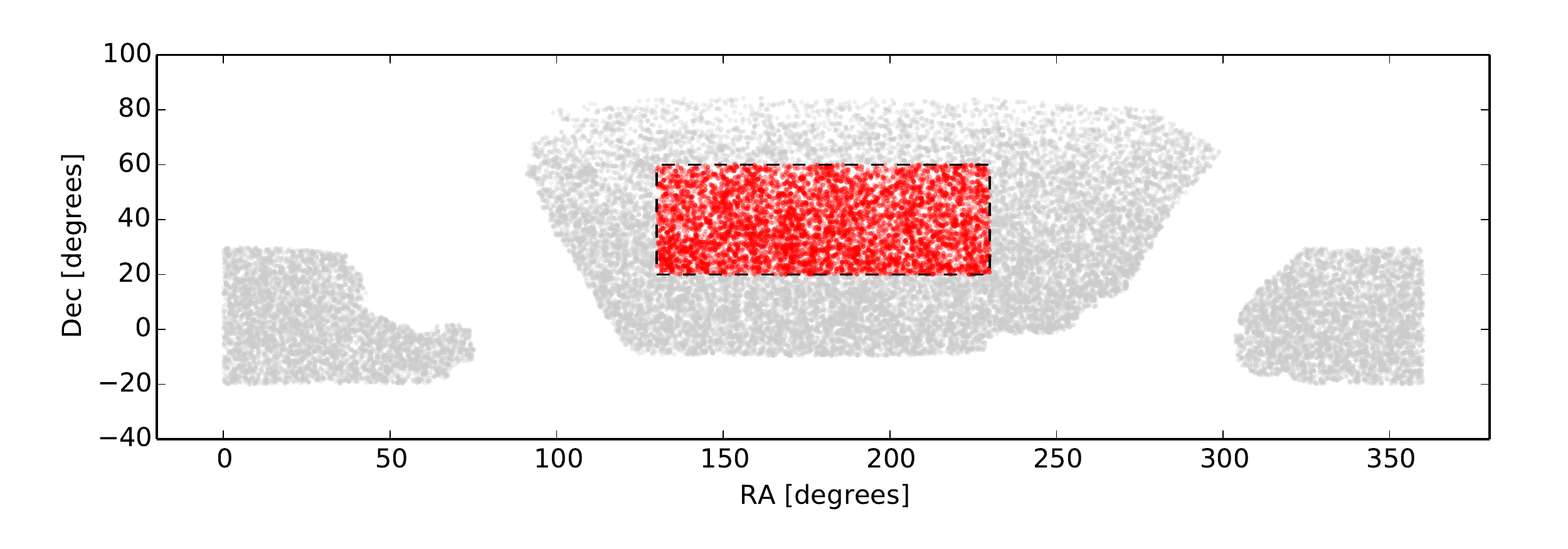}
   \caption{Expected DESI footprint (grey) versus the subset considered in
     this work (red): only a subset is considered to speed up the
     calculation of the correlation functions. The volume ratio between the two samples is $\approx 0.19$.}
  \label{fig footprint}
 \end{center}
\end{figure*}

In this work we consider mocks with an area of $\sim 3,000$ deg$^2$
containing samples of the DESI dark-time targets: Luminous
  Red Galaxies (LRGs), Emission Line Galaxies (ELGs) and Quasars
  (QSOs), with redshift distributions matching those expected for
  DESI, as given in figures 3.8, 3.12, 3.17 of \citet{amir2016a},
  respectively.  As discussed in section 4.2 of the same paper, the
total footprint for DESI will be $\sim 14,000$ deg$^2$, i.e. about 5
times larger than that of our mocks (see Fig.~\ref{fig footprint}).
Since the tiling does not depend on density and we are not interested
in the absolute error determined from the survey, we are allowed to
use any sized mock as long as we have enough statistical precision on
the scales of interest.  As these scales are quite small, 3,000
deg$^2$ is sufficient.

We call the pointings of the instrument on the sky the tiling, and the allocation
of fibres to individual galaxies depends on a fibre-allocation
algorithm. These are expected to be independent, with the density of
targets not affecting the tiling. This simplifies the analysis, as the
effect of the fibre assignment can be determined at fixed tiling.

\begin{figure*}
 \begin{center}
   \includegraphics[width=7.5cm]{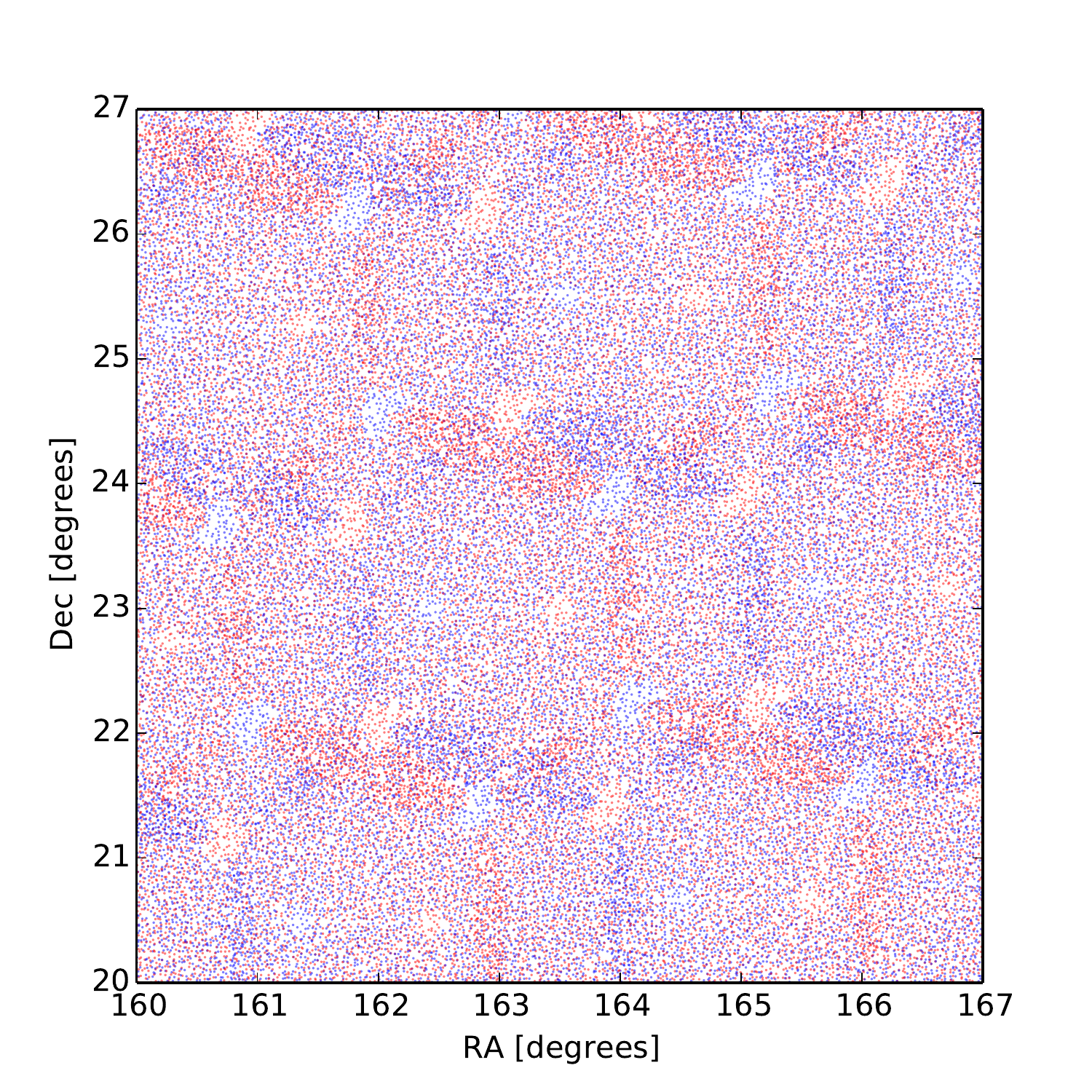}
   \includegraphics[width=7.5cm]{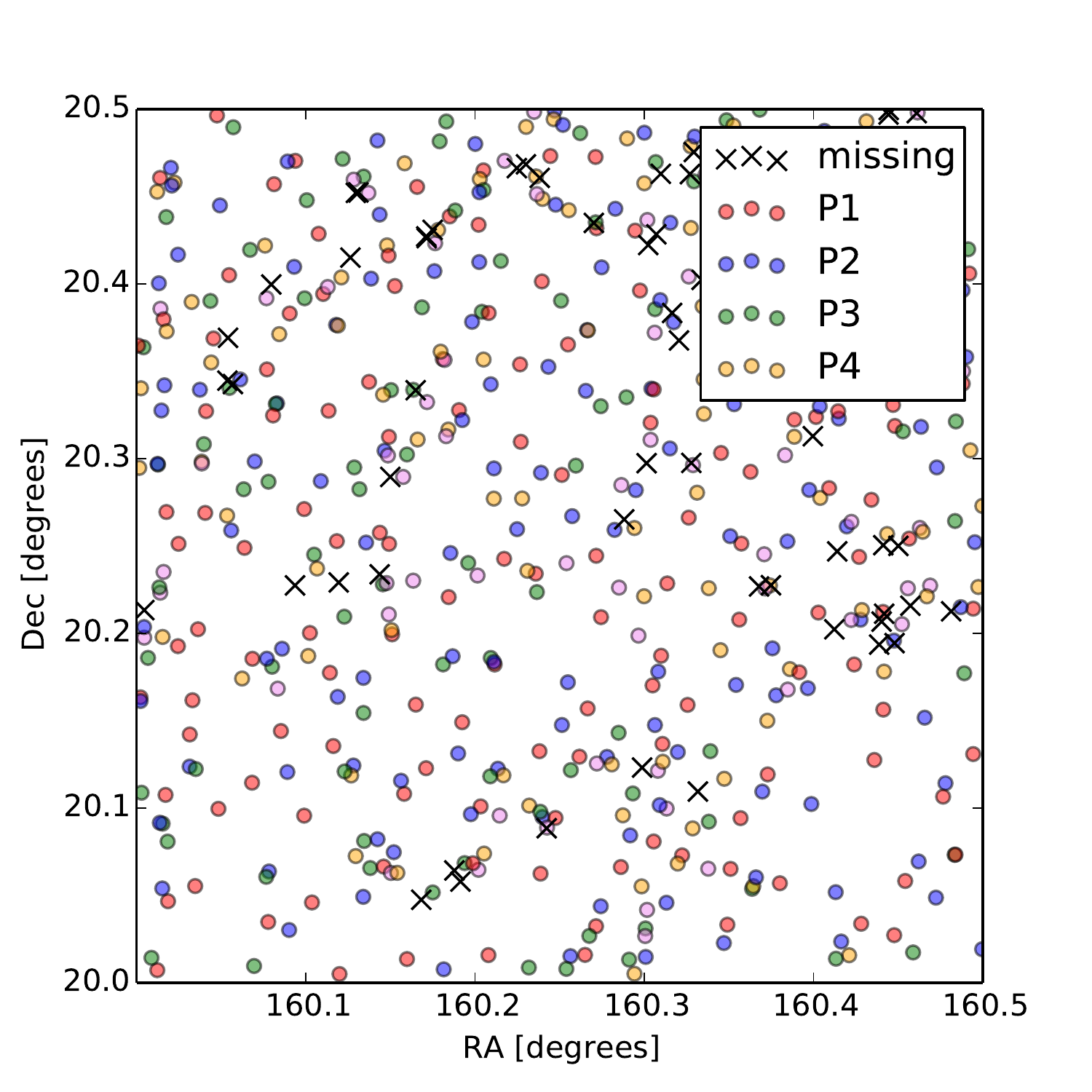}   
   \caption{Close-ups of the observing strategy for two different levels of zoom.
     Different colour dots show observations made in different passes, whereas crosses indicate galaxies that are never observed.
     The left panel shows the imprint of the focal-plane pattern on the sky  (only the first two passes are reported, for clarity).
     In the right panel we focus on a smaller area in order to emphasise how the completeness depends on the density: on average, in crowded regions the number of missing observation is larger and, more in general, galaxies in these regions cannot be observed simultaneously, i.e. circles too close one each other have different colours.}
   \label{fig Obs closeup}
 \end{center}
\end{figure*}

The fibre positioner locations for the DESI instrument are shown in
figure~4.2 of \citet{amir2016a}. The circular focal plan is split
into 10 petals, each of which holds 500 fibre positioners. There are
areas with no coverage at the centre, between petals, regions removed
for guide focus arrays, and for positioners set as guides for the
fibre view camera. In addition, the tiling of the survey region allows
for overlap of neighbouring pointings. The resulting coverage pattern
is complicated and the probability of a target galaxy being selected
for observation also varies with local density, giving rise to non-cosmological density dependent fluctuations
on a number of scales. Even after multiple passes, the imprint of the
instrument on the sky is clearly visible, as shown in Fig.~\ref{fig Obs closeup}.

The largest galaxy survey undertaken by DESI will be of ELGs, with an expected total number of $\sim$18M, spanning
a redshift range out to $z\sim1.6$. The parent sample of targets is selected from
imaging data and fed into the fibre-assignment
algorithm, which calculates the correct placement of the fibres to
match targets.
ELGs compete for fibre assignment with two other classes of targets: LRGs and QSOs.
Consequently, the number of ELGs with no fibre assigned depends on the spatial distribution of all three galaxy types.
The expected cumulative completeness after a number of passes for
the ELG sample is given in the upper row of Table~\ref{tab stats}, where we introduce the obvious notation P1,...,P5 to indicate different number of passes of the instrument.
\begin{table}
 \centering
  \begin{tabular}{rllllll}
   & P1 & P2 & P3 & P4 & P5 \\
  \hline
  \hline
  full mocks [\%] & $23$ & $44$ & $61$ & $73$ & $81$ \\  
  ELG mocks [\%] & $25$ & $48$ & $67$ & $81$ & $90$
  \end{tabular}
  \caption{The cumulative percentage of emission line targets that are observed after a certain number of passes. 
  Upper row: ELG completeness when the fibre-assignment algorithm is applied to mocks that include all possible DESI targets, assuming the current priority scheme.
  Lower row: ELG completeness when only ELGs are targeted.
  In both cases, the reduction of the efficiency with increasing number of passes can clearly be seen: the number of redshifts measured with pass five alone is about 3 times smaller than that measured with pass one.}
 \label{tab stats}
\end{table}

In this work we use a preliminary version of the fibre-assignment algorithm, namely \texttt{fiberassign 0.2.3}, which is available at the following link: \url{https://github.com/desihub/fiberassign/releases/tag/0.2.3}.

\section{Model}\label{sec model}

In this section we provide a compact description of the weighting
scheme adopted throughout the paper. For a more detailed discussion
see \citet{bianchi2017} and \citet{percival2017}, in which this
technique was originally introduced.  

At each separation $\vec{s}$, we compute the two-point correlation function
$\xi$ via the minimum-variance estimator \citep{landy1993},
\begin{equation}\label{eq ls}
  \xi(\vec{s}) = \frac{DD(\vec{s})}{RR(\vec{s})} - 2 \frac{DR({\vec{s}})}{RR(\vec{s})} + 1 \ ,
\end{equation}
where, as usual, $DD$, $RR$ and $DR$ stand for data-data,
random-random and data-random pairs, respectively.  Eq. (\ref{eq ls})
is supplemented with the pairwise-inverse-probabilty (PIP) weights
defined in \citet{bianchi2017},
\begin{equation}\label{eq PIPauw}
  DD(\vec{s}) = \sum_{\vec{x}_m - \vec{x}_n \approx \vec{s}} w_{mn} \frac{DD^{(p)}_a(\theta)}{DD_a(\theta)} \ ,
\end{equation}
where $w_{mn}=1/p_{mn}$ is the inverse selection probability of the
pair formed by the galaxies $m$ and $n$, whereas $DD^{(p)}_a$ and
$DD_a$ represent the angular pair counts of the parent and observed
sample, respectively\footnote{The parent sample is the set of all possible targets, for which we have imaging data, i.e. angular positions. The observed sample is the subset of galaxies that have been selected by the fibre-assignment algorithm, i. e. the galaxies for which we actually have a spectroscopic redshift.}.
This latter is, in turn, computed via the same $w_{mn}$ weights,
\begin{equation}\label{eq PIPauw ang}
  DD_a(\theta) = \sum_{\vec{u}_m \cdot \vec{u}_n \approx \cos(\theta)} w_{mn} \ .
\end{equation}
The symbols $ \sum_{\vec{x}_m - \vec{x}_n \approx \vec{s}}$ and
$\sum_{\vec{u}_m \cdot \vec{u}_n \approx \cos(\theta)}$, with
$\vec{u}_i = \vec{x}_i/|\vec{x}_i|$, indicate that the sum is
performed in bins of $\vec{s}$ and $\theta$, respectively.  Similarly,
with obvious notation,
\begin{equation}\label{eq PIPauw DR}
DR(\vec{s}) = \sum_{\vec{x}_m - \vec{y}_n \approx \vec{s}} w_m \frac{DR^{(p)}_a(\theta)}{DR_a(\theta)} \ ,
\end{equation}
where $w_m=1/p_m$ is the inverse selection probability of the galaxy
$m$, and
\begin{equation}\label{eq PIPauw DR ang}
  DR_a(\theta) = \sum_{\vec{u}_m \cdot \vec{v}_n \approx \cos(\theta)} w_m \ .
\end{equation}

To evaluate the pair weights, $w_{mn}$, we adopt the following procedure.
The targeting outcomes for each galaxy (when the fibre-assignment algorithm is run $N_{bits}$ times) are stored as bits where 1 represents galaxy selection in that realisation and 0 when the galaxy is discarded. 
The bitwise weight of a galaxy $w_i^{(b)}$ is then a binary array of length $N_{bits}$, which, for convenience, we encode in base-ten integers.
Once the galaxies are assigned these weights, pair weights are computed ``on the fly'' while doing pair counts by comparing the two bitwise arrays of each galaxy. 
The pair weights are the total number of realisations divided by the number of realisations in which both galaxies are selected and therefore would contribute to the pair counts,
\begin{equation}\label{eq wb DD}
  w_{mn} = \frac{N_{bits}}{\popcnt\left[w^{(b)}_m \ \& \ w^{(b)}_n \right]} \ ,
\end{equation}
where $\&$ and $\popcnt$ are fast bitwise operators, which,
respectively, (i) multiplies two integers bit by bit, and (ii) returns
the sum of the bits of the resulting integer.  Similarly, for
individual weights we have
\begin{equation}\label{eq wb DR}
  w_m = \frac{N_{bits}}{\popcnt\left[w^{(b)}_m \right]} \ .
\end{equation}
One obvious advantage of adopting this ``on-the-fly'' approach is that, by doing so, we do not have to to deal with the problem of storing $\sim 30M \times 30M$ real numbers (pair weights). 

In general, when there are no zero-probability pairs, the PIP approach is unbiased by construction and the only effect of angular upweighting is to reduce the variance of the weights.
Unfortunately, the non-zero-probability condition is formally violated for the fibre-assignment algorithm considered in this work, as discussed in Sec. \ref{sec zero prob}.
Thus we need to include angular upweighting in the analysis not only to improve the variance but also to ensure unbiased $\xi$ measurements (see Sec. \ref{sec zero prob}).

\section{Dealing with selection probabilities}\label{sec zero prob}

Given a volume $V$ from which we want to recover the clustering, in order to be unbiased, the PIP correction requires that no pair in $V$ has zero probability of being observed.
The DESI fibre-assignment algorithm takes as input the tiling positions and then runs a stochastic process, initialised by random seeds, to select which galaxies will be observed\footnote{In its most recent version, the algorithm is not stochastic anymore, i.e. there are no random seeds. Instead, the set of galaxies to be observed is now specified by a subpriority list that the algorithm takes as an input. This just moves the stochasticity from the algorithm to the list and does not affect our conclusions.}.
If we assume fixed tiling positions and apply the algorithm to the volume defined by the footprints in Figure \ref{fig footprint}, either the full DESI area or the subset considered in this work, the non-zero-probability condition is violated, especially after a single pass of the instrument.

One obvious example of pairs that cannot be observed with a single pass is (i) the set formed by galaxies that fall where the guide focus arrays are (see the gaps in the instrument pattern in Figure \ref{fig Obs closeup}, left panel, red dots only).
Another important zero-probability set is that of (ii) the close pairs not belonging to overlap regions, for which, due to the impracticability of allocating two fibres at a separation smaller than their physical size, it is impossible to measure simultaneously the two redshifts.

There are different possible countermeasures for these effects, (i) can be solved, e.g., by redefining $V$, which, in practice, means to put holes in the random sample in correspondence of the guide focus arrays (and similarly for all the other blind spots in the focal plane).      
Adjusting the random sample would not work perfectly for (ii), due to the entanglement between clustering and fibre-allocation issues.   

On the other hand, we note that both (i) and (ii) can be considered just tiling-related issues.
Since the choice of the tiling locations does not depend on the clustering, (i) and (ii) can be solved at once by either (a) allowing for rigid displacements of the tiling when evaluating the probabilities (the choice of the tiling is itself a random choice) or (b) relying on the angular upweighting to restore the correct pair counts while keeping the tiling fixed.
The philosophy behind (b) can be roughly summarised as ``if the unobservable pairs have nothing special then their clustering can be deduced by the pairs we observe''.

For this work we adopt  strategy (b), but either options would be able to correct for the fibre assignment.
The only advantage of one method over the other would be to make the weights more even and reduce the shot noise in the weighted catalogue.
More specifically, approach (b) formally makes the PIP weights less even, so it is possible that the errors that we show in Section \ref{sec results} can be further reduced by adopting approach (a). On the other hand, approach (a) would require more realisations of the targeting for a fair sampling of the probability.

When objects other than ELGs are considered, the presence of zero-probability pairs is not only a consequence of tilling but also of higher priority objects taking precedence.
Problems occur when there are correlations between higher priority populations and the ELG population.
In this work we run the fibre-assignment algorithm directly on ELGs, so the problem is ignored, but it is important to be aware that a too strict priority scheme is a potential cause of systematics.
It is also important to note that this issue is not created but rather made apparent by the inverse-probability weighting.
It is a general fact that if we decide a priori not to observe a subset of a galaxy population of interest then we must be sure that we understand the behaviour of the missing galaxies, e.g. as when the populations are uncorrelated and, as a consequence, the effect on lower priority targets is just a random dilution.  

A simple way around this obstacle would be, e.g., to reduce the probability of picking a quasar from 100\% to 90\% of the times when an allocation conflict with some lower priority target happens, and similarly for all classes of galaxies.
Alternatively, we could reobserve small subsets of the whole survey area until 100\% completeness is reached for all the populations of interest.     
Both options would ensure that pairs containing all types of target were observed at some part of the survey.

\section{Mock catalogues}\label{sec mocks}

For our analysis we use the same set of QPM mock catalogues described
in \citet{burden2017}.  Briefly, we perform our analysis on 25 Quick
Particle Mesh (QPM, \citealt{white2014}) mock galaxy catalogues
designed specifically to mimic the DESI survey. Dark matter
distributions were generated using a low resolution particle mesh
N-body code, based on initial conditions set at z=25 using second
order Lagrangian perturbation theory. At each step a subset of
particles (chosen based on local density) are selected as dark matter
halos and assigned a halo mass. The mass values are tuned using higher
resolution simulations. Haloes are populated with galaxies using the
functional form of the Halo Occupation Distribution function of
\citet{tinker2012}. Although we know that this form has difficulty
fitting the expected ELG distribution \citep{Gonzalez-perez2017}, the
work presented here does not depend on the exact form of the true
clustering signal.

These mock catalogues were sampled using the current baseline
algorithms  \citep{cahn2015} to mimic the DESI data. As the exact targeting, tiling and
fibre allocation algorithms for DESI may still be further optimised
before the survey commences, these mocks will almost certainly be
different in detail from the final survey. However, the validity of
these mocks is not important: all we require is that they are affected
by the same systematic effects as the future data, so we can show that
we can remove those effects.

The mocks include all of the classes of galaxies to be observed in the
DESI wide survey, undertaken in dark time. The classes of objects are,
in the order of the priority with which they are observed:
Lyman-$\alpha$ QSOs, QSOs, LRGs and ELGs. So, although ELGs are the
most numerous galaxy sample they are the lowest priority, and
consequently, even after five passes, we only expect $\sim$80\%
completeness, and incompleteness will remain a significant effect to
be corrected. For this reason, in this paper we concentrate on this
sample, although the methodology is also applicable to the QSO and LRG
samples.
In general, we expect the fibre-assignment issue to be less severe for these latter classes of targets, not only because of their higher priority but also for their lower surface density, see, e.g., table 3.1 in \citet{amir2016a}.

\section{Recovered clustering measurements} \label{sec results}

\subsection{The correlation function}\label{sec results: xi}

We measure the clustering (using the first three even Legendre multipoles) of four realisations of the fibre-assignment algorithm for each of the 25 mocks.
Since the expected final ELG completeness is reached after four passes of the DESI instrument when ELG-only mocks are considered (Table \ref{tab stats}), we limit our analysis to the first four stages of the observing strategy.
In total we measure the correlation function moments for 400 different mocks of DESI data.

We run the fibre-assignment algorithm for each mock catalogue
992 times\footnote{We store the bitwise weights in 32-bit integers, but, for simplicity, we only consider positive integers.
This means that we actually use only 31 bits per integer.
It is therefore convenient to use a number $N_{bits}$ of targeting realisations that is a multiple of 31, in our case $N_{bits}= 992 = 31 \times 32 \approx 1000$.} in order to calculate the PIP weights\footnote{Although we run the selection algorithm 992 times, we only measure the clustering for four of the 992 realisation, in order to save computational time.
By using all the available realisations, due to exact cancellations, we would have a better behaved mean of the clustering but similar error bars (see \citealt{bianchi2017}).}.
As discussed in Section \ref{sec zero prob}, for each run of the fibre selection algorithm, we change the random number used to select
targets to be assigned fibres, but do not move the location of the
DESI survey, i.e. the probability we use is that given the fixed tiling locations chosen.

\begin{figure*}
 \begin{center}
   \includegraphics[width=5.7cm]{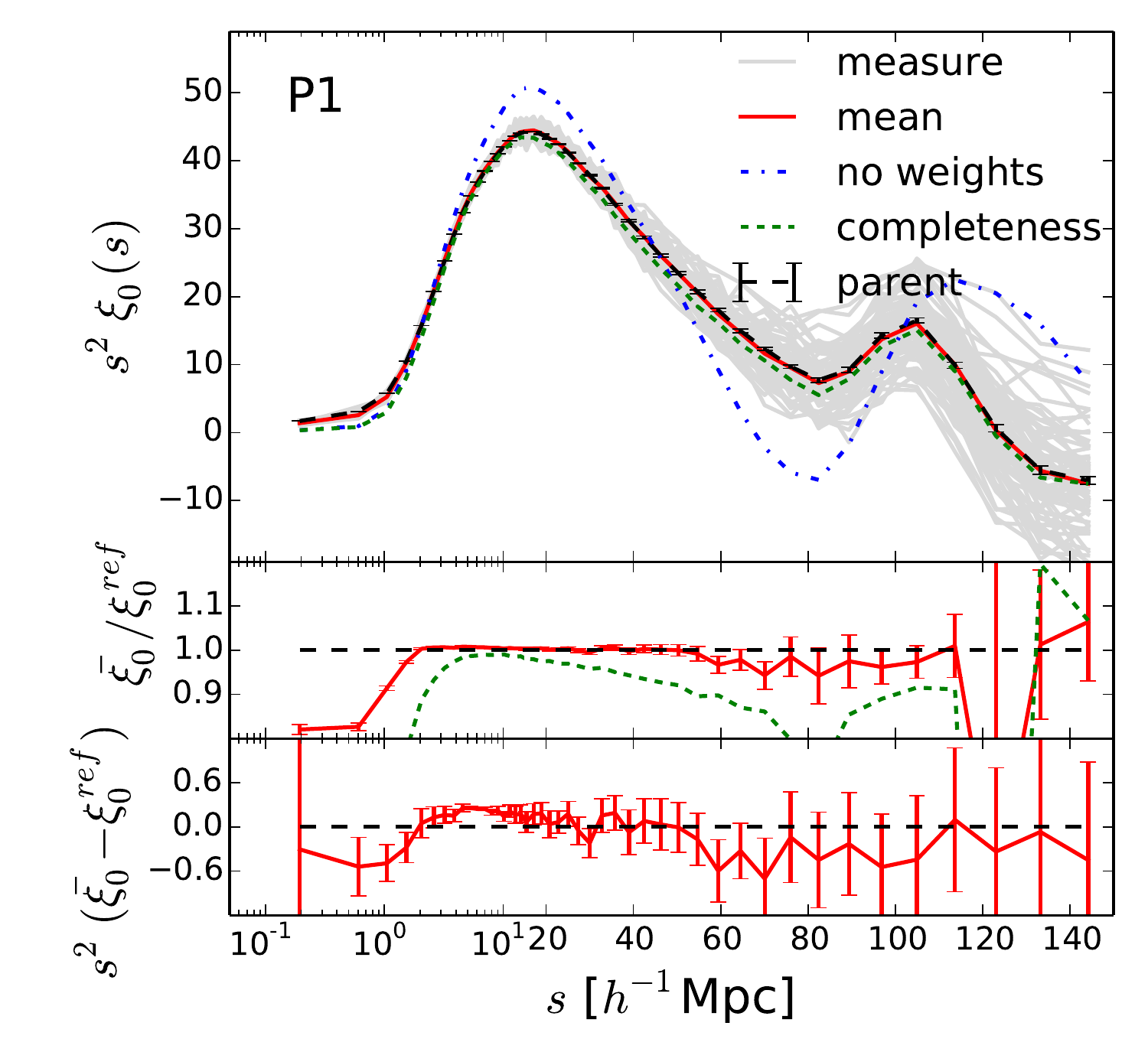}
   \includegraphics[width=5.7cm]{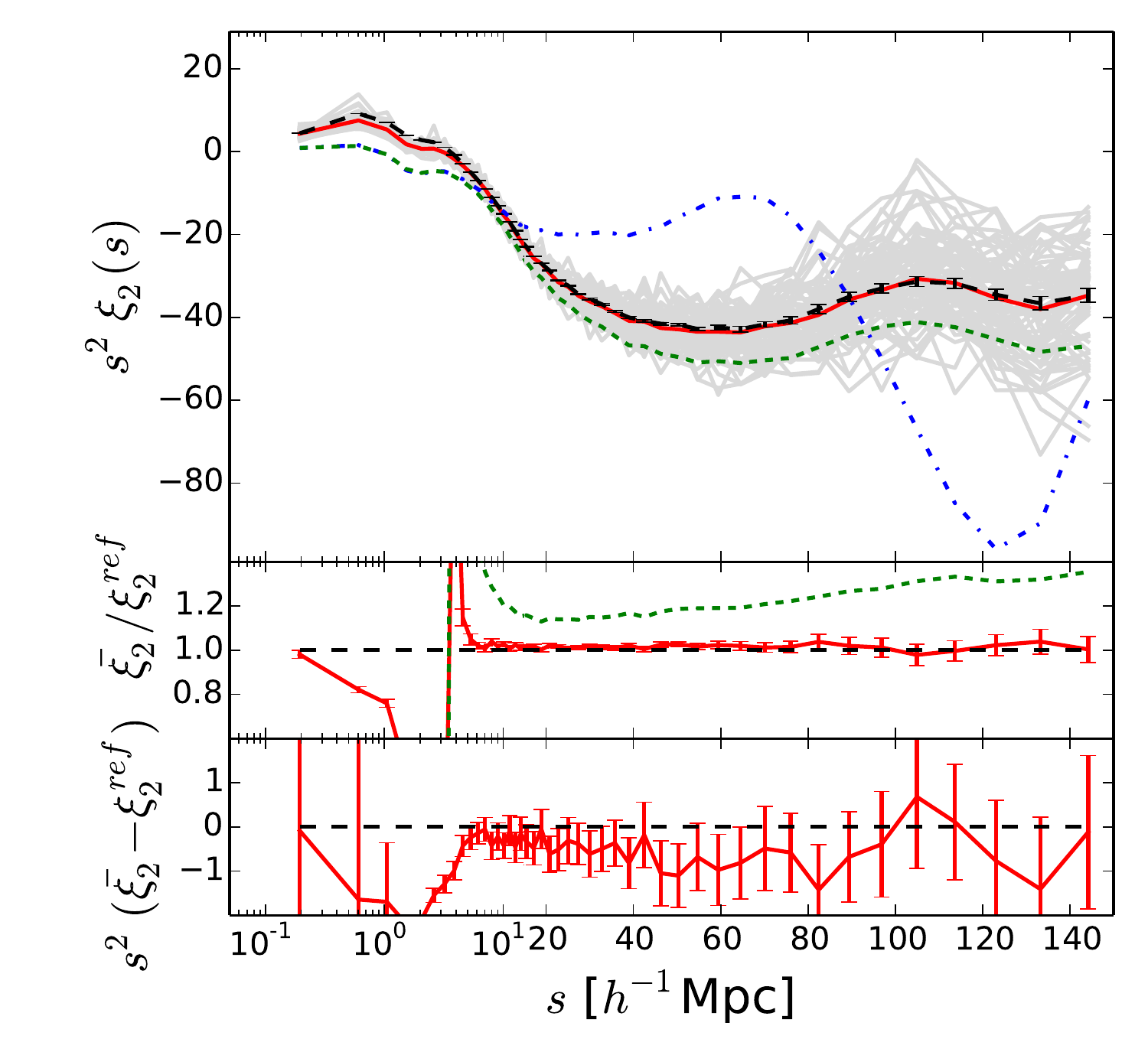}
   \includegraphics[width=5.7cm]{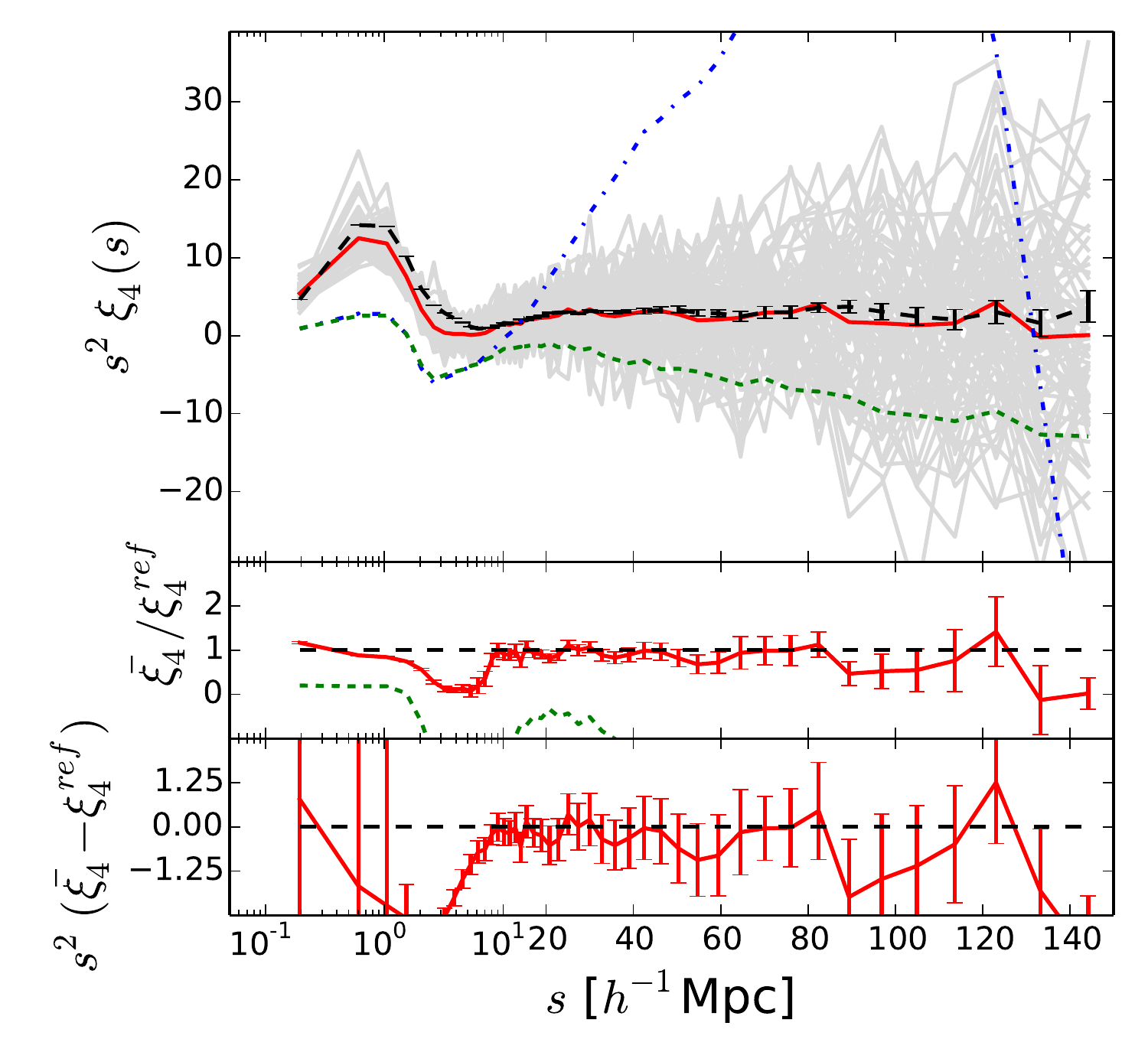}\\
   \includegraphics[width=5.7cm]{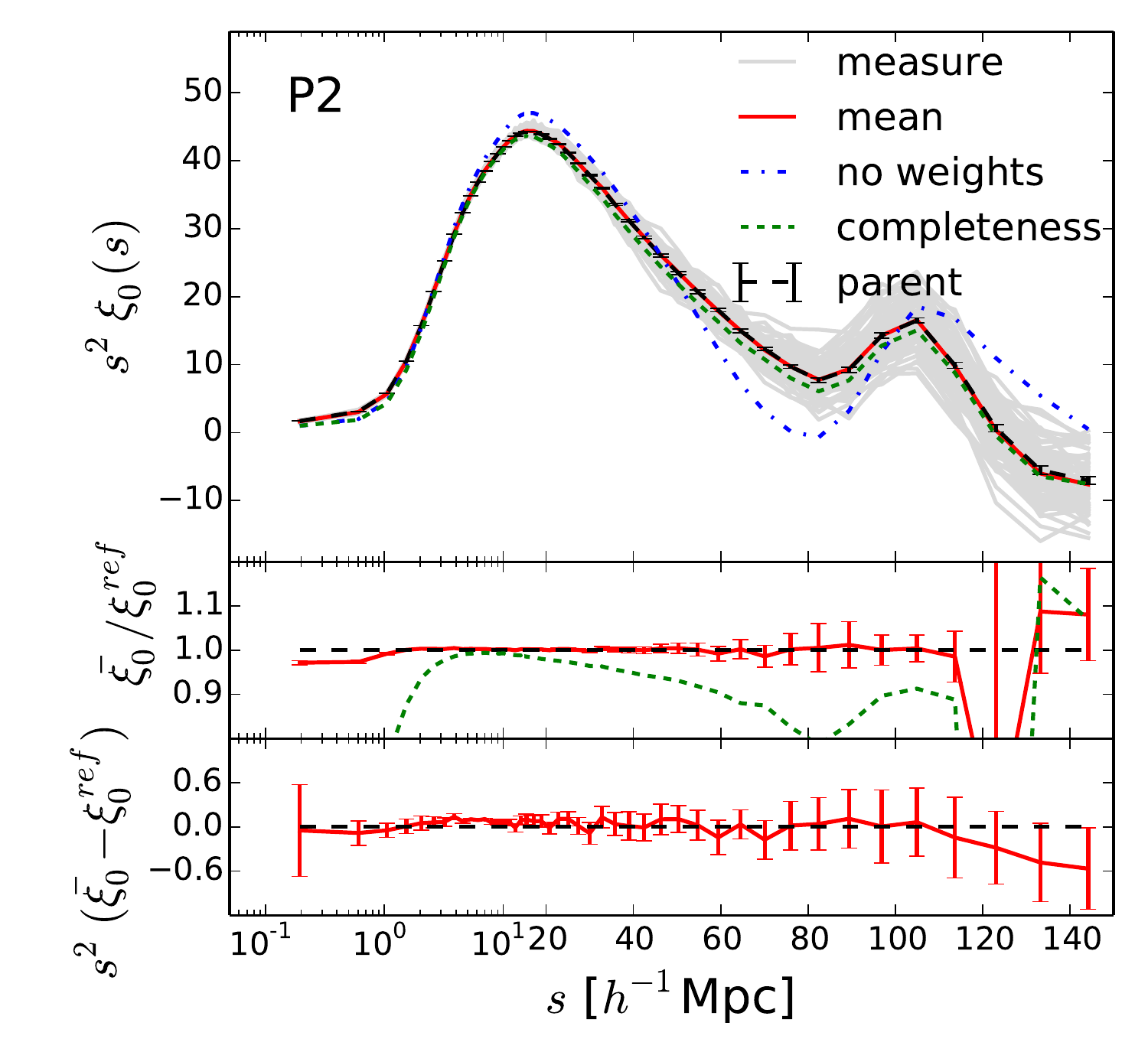}
   \includegraphics[width=5.7cm]{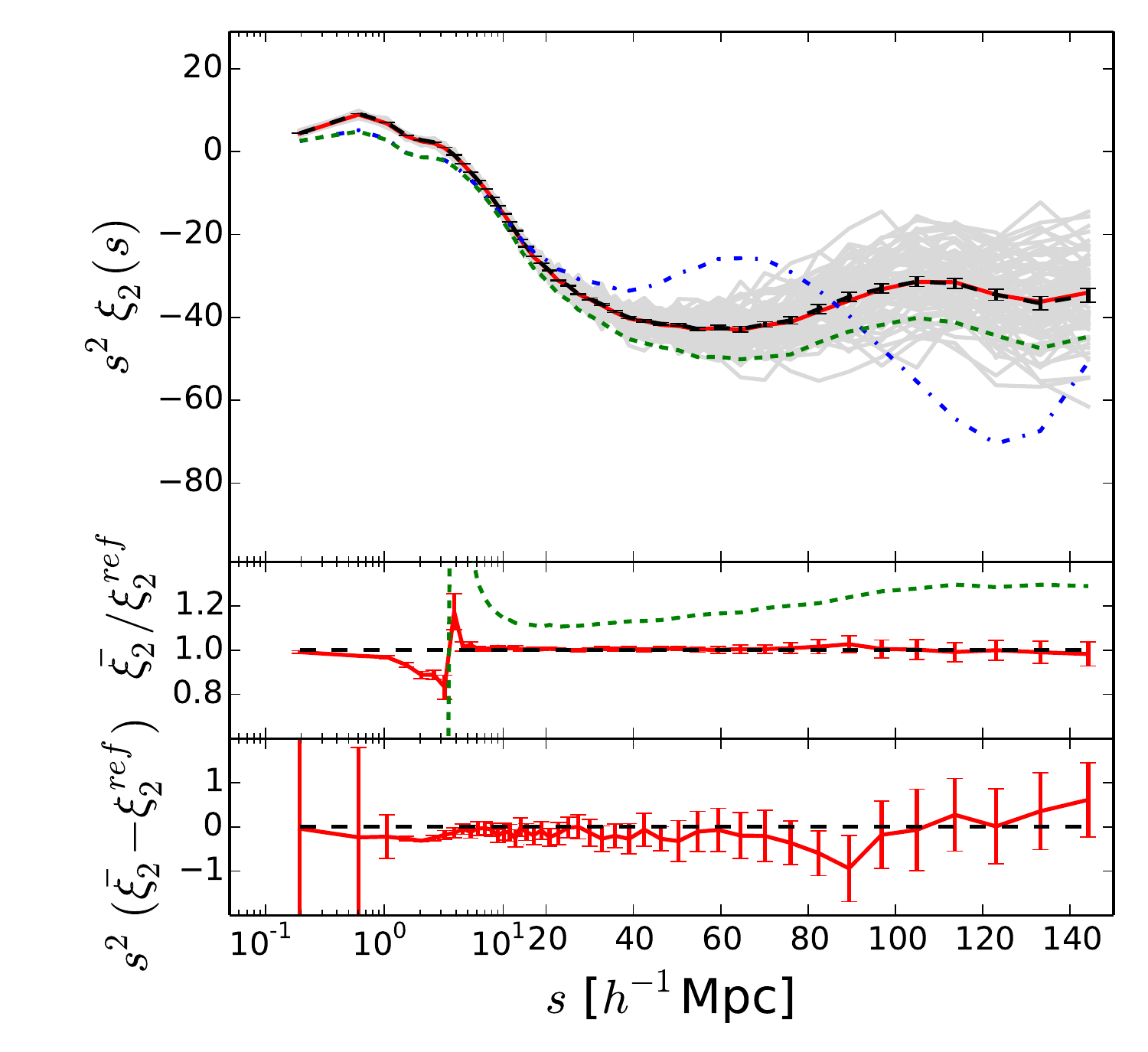}
   \includegraphics[width=5.7cm]{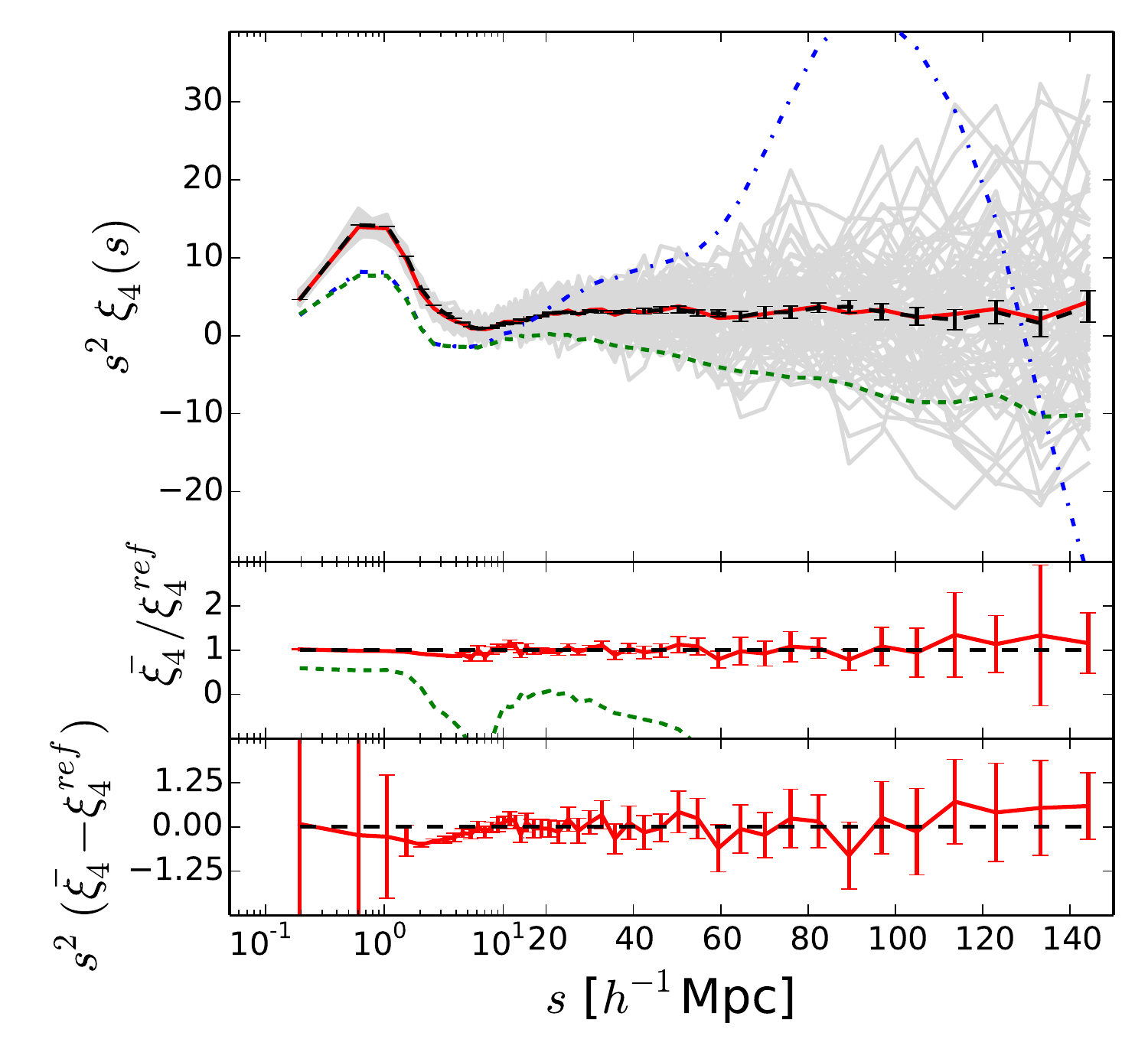}\\
   \includegraphics[width=5.7cm]{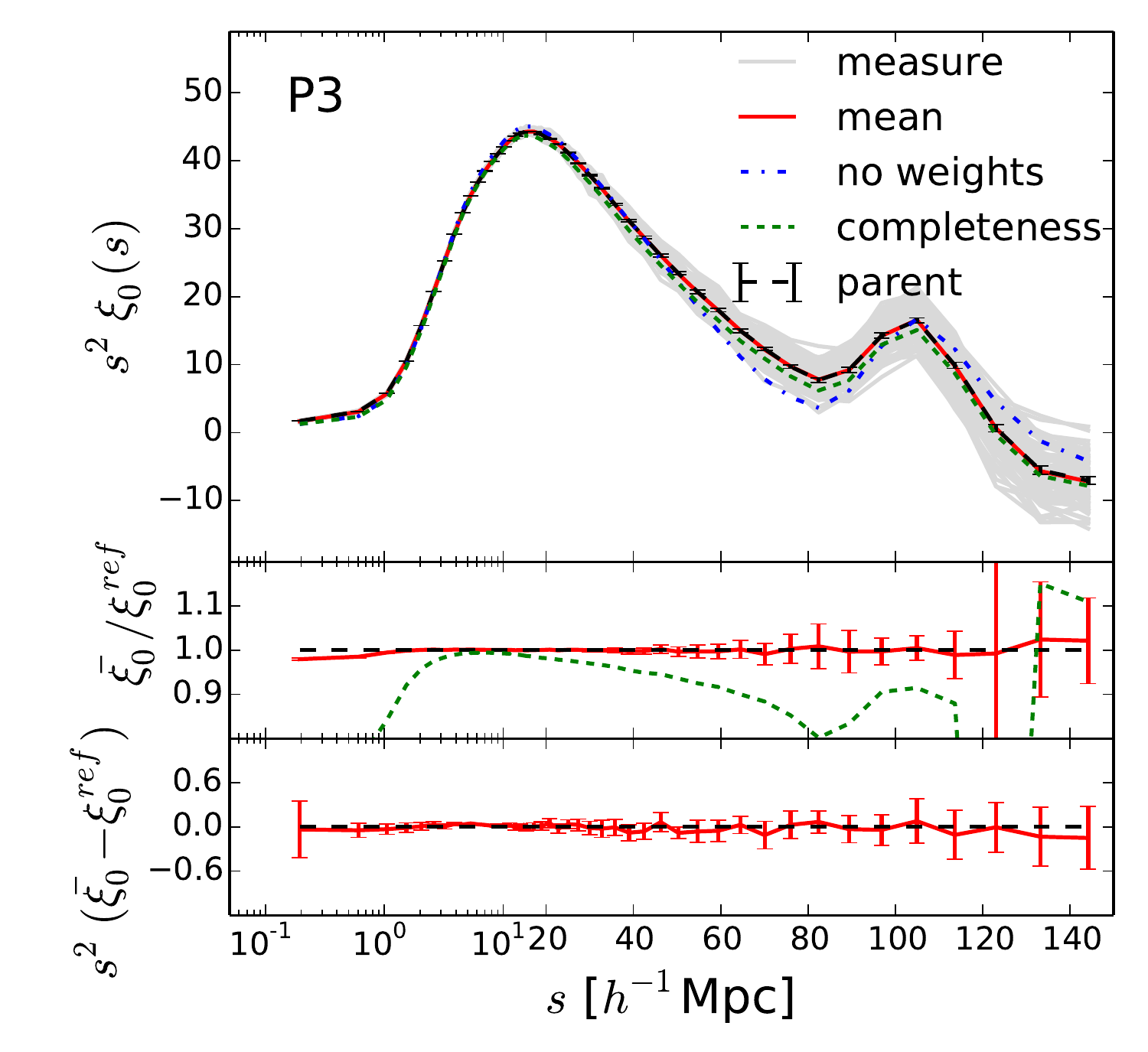}
   \includegraphics[width=5.7cm]{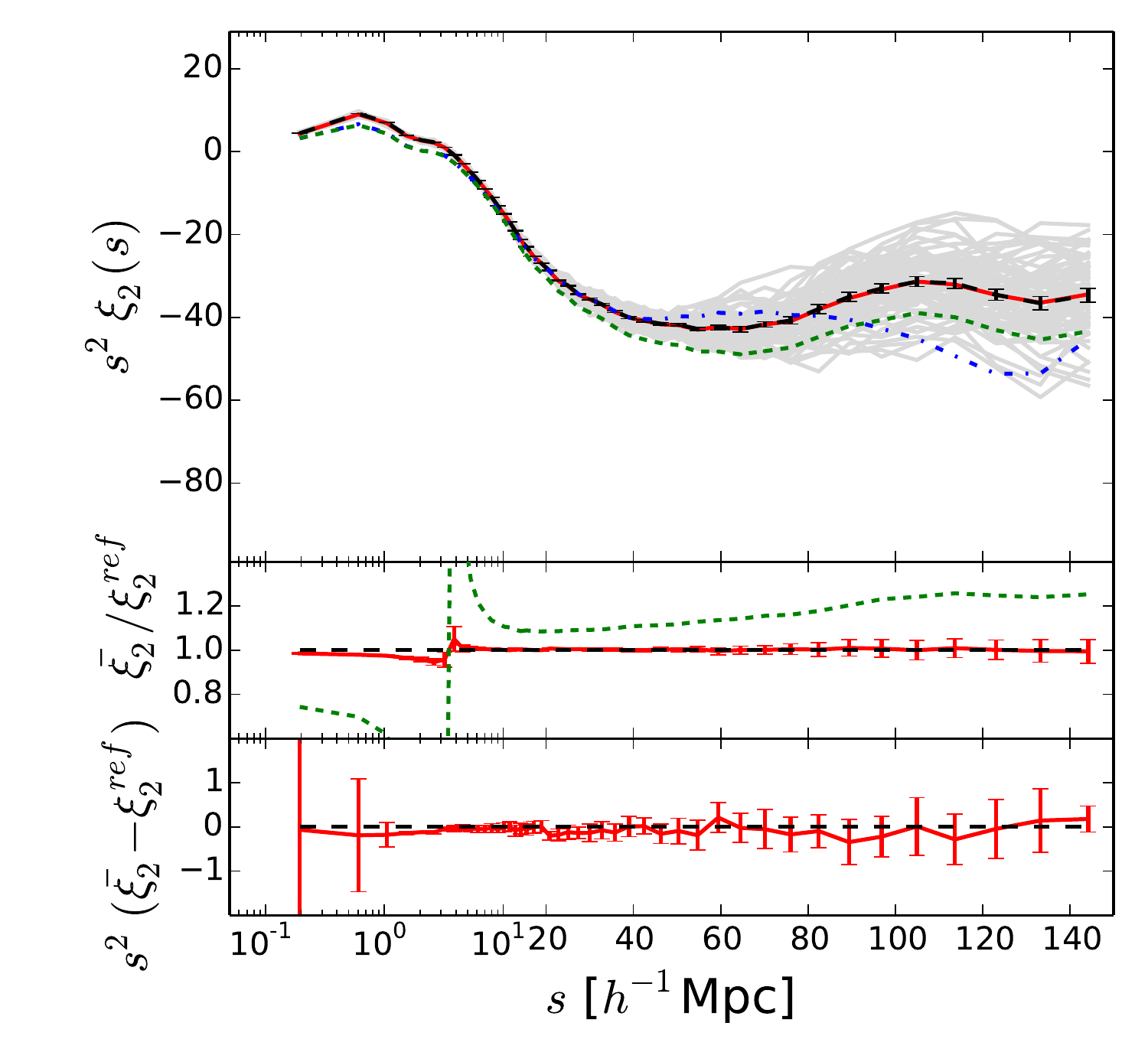}
   \includegraphics[width=5.7cm]{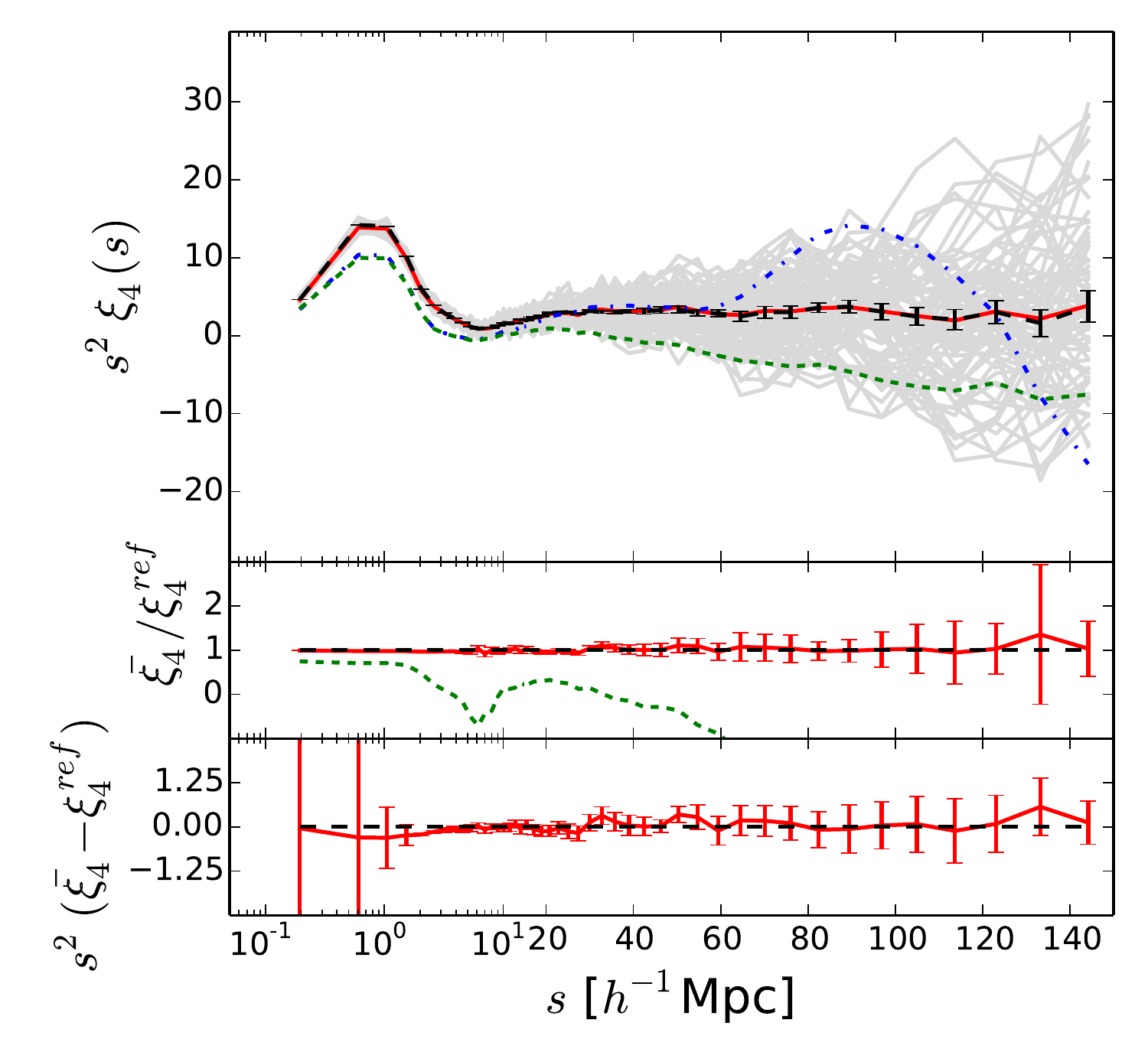}\\
   \includegraphics[width=5.7cm]{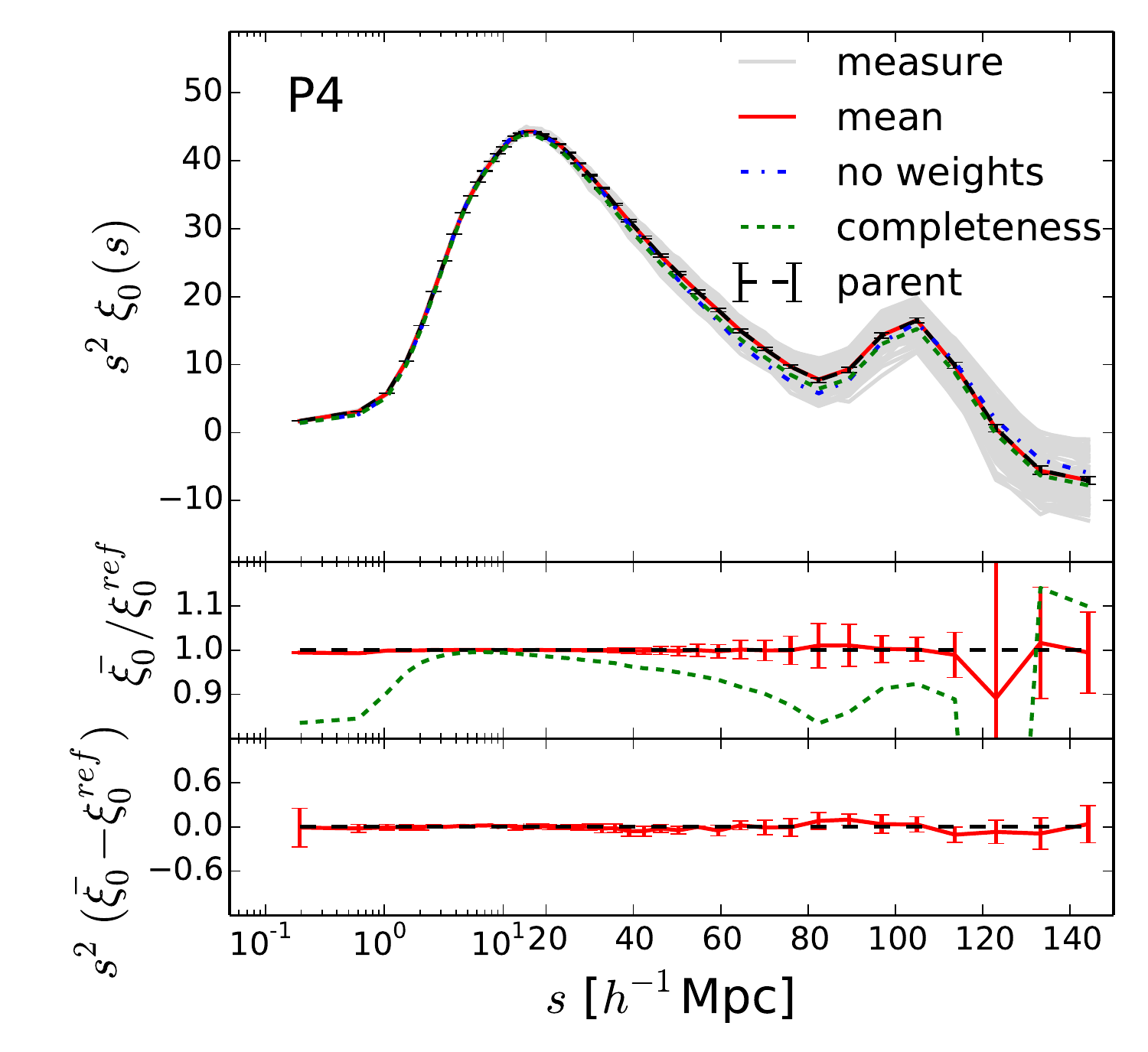}
   \includegraphics[width=5.7cm]{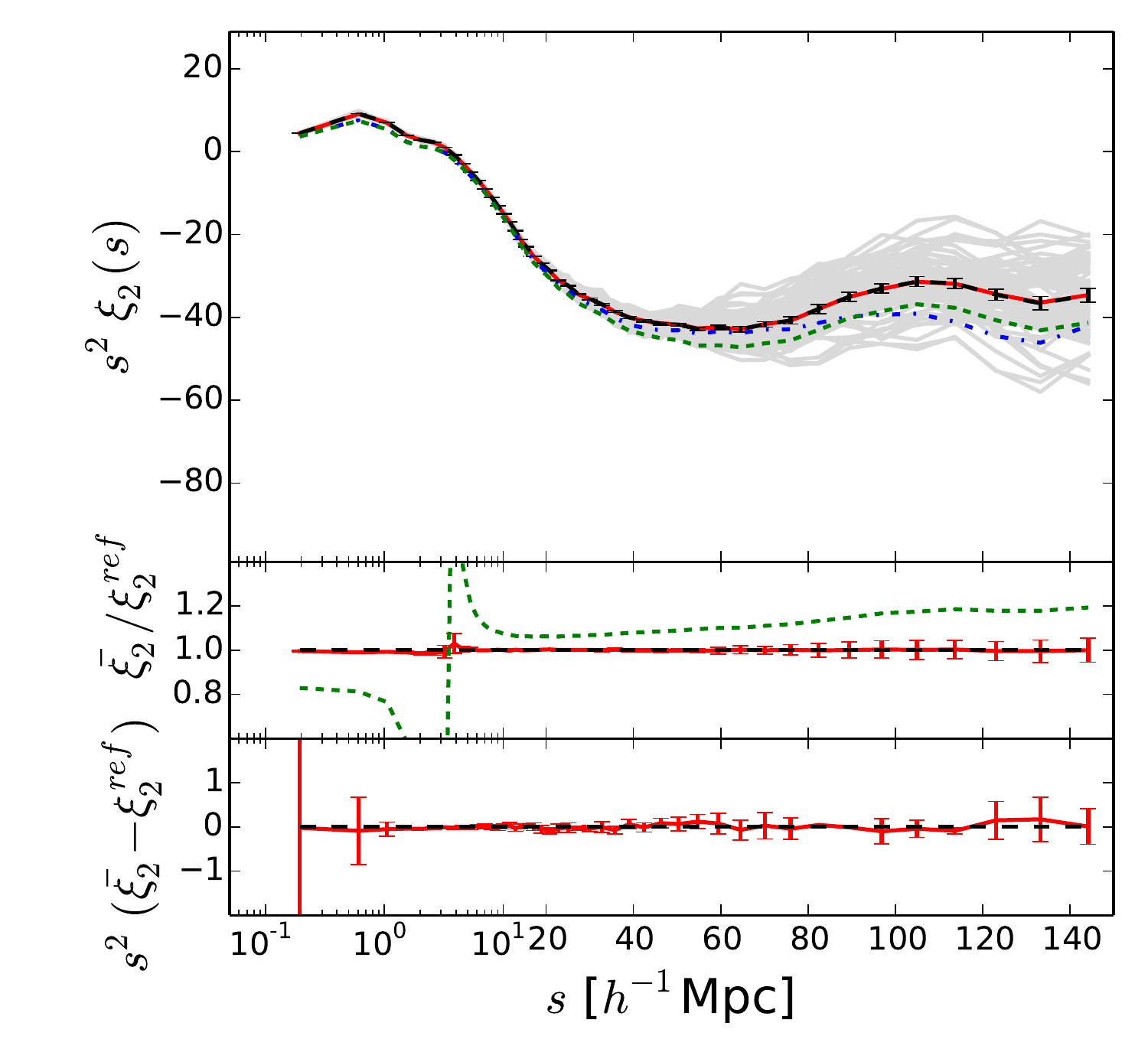}
   \includegraphics[width=5.7cm]{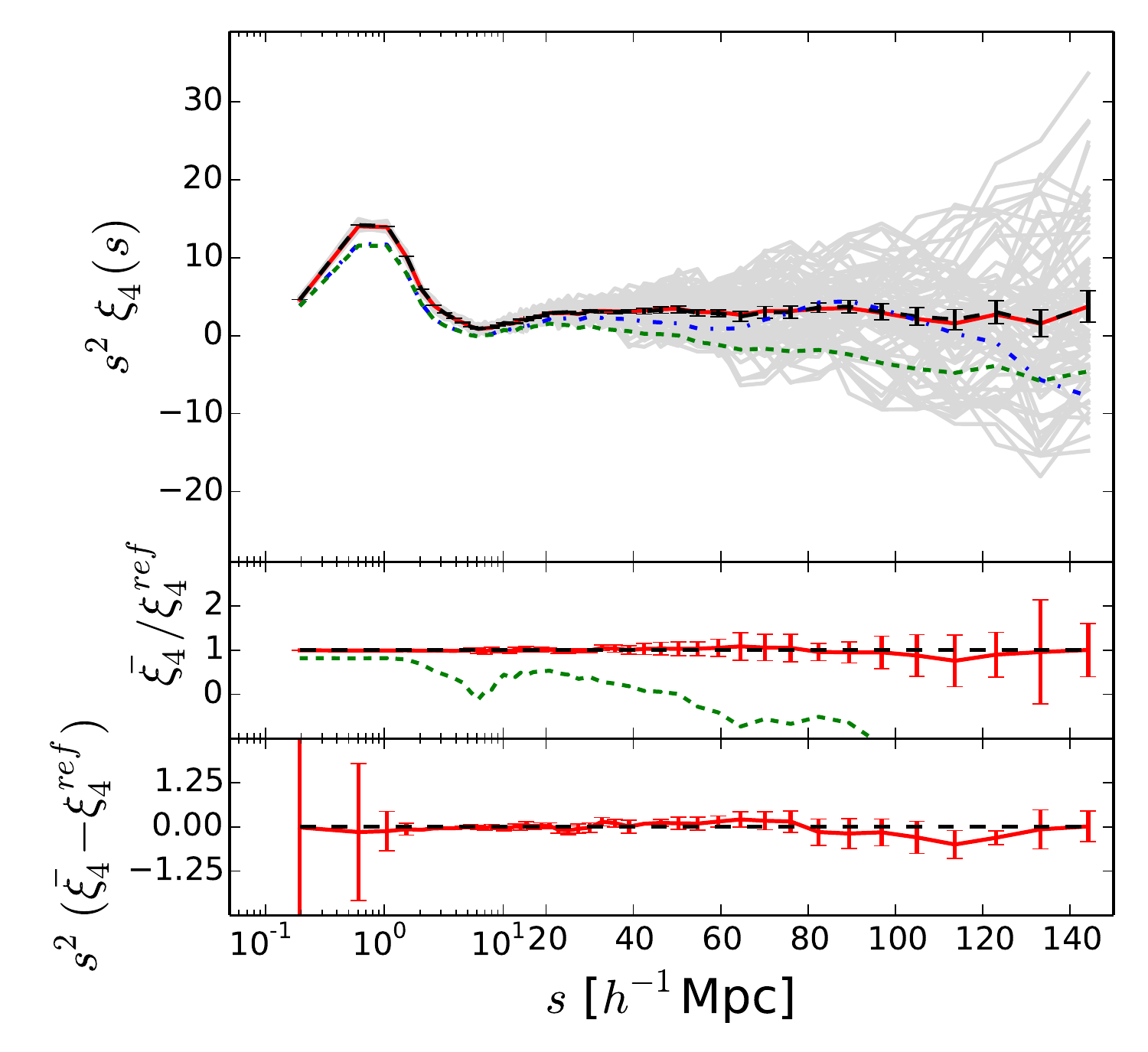}
   \caption{The measured multipoles of the correlation functions from the DESI mocks,
     after correcting for partial selection of the targets using the
     PIP weights (solid grey lines) compared to their mean (solid red lines) and for the full parent sample
     (black dashed lines).
     In order to visualise both large- and small-scale features at once, we adopt a logarithmic scale for the abscissa for $s_\perp < 15 h^{-1}$Mpc and linear elsewhere.     
     The uncorrected multipoles are shown by the dot-dashed blue lines. Strong deviations from the true clustering signal are observed.
     These deviations get smaller when a completeness map is applied to the random sample, green dashed.
     Still,  the expected value is not recovered, with discrepancies that grow with the order of the multipoles.}
   \label{fig vamo}
 \end{center}
\end{figure*}

In Fig.~\ref{fig vamo} we present the measured Legendre polynomial
moments of the correlation function: panels are split by the order of
the moment (horizontal) and by the number of passes (vertical).
See Appendix \ref{sec details} for technical details on how we compute these quantities. 
The mean of the measurements corrected using the PIP weights is compared
to the reference value given by the mean of the multipoles measured
from the same mock catalogues when no subsampling by the fibre
assignment algorithm is applied. In the bottom frame of each panel we
report the ratio between the mean of the measurements and the
reference value, with error bars obtained by propagating the error on
both these two quantities. Agreement within the expected errors is
shown.

The blue dot-dashed lines show the multipole moments calculated
without the PIP weighting: clearly here we see a large difference
between corrected and uncorrected measurements, particularly for a low
number of passes.

The performance of the unweighted measurements can be improved by removing from the clustering signal density fluctuations caused by blind spots in the focal plane and tile-overlaps, i.e. the (clustering-independent) contributions purely coming from the imprint of the DESI instrument and tiling.
This can be obtained by mimicking such fluctuations with the random sample.
We therefore create a clustering-independent completeness map, $c(\text{RA}, \text{Dec})$ (see App. \ref{sec comp map}), and impose it to random sample.
The corresponding measurement are shown in dotted green.
Compared to the previous correlation function measurements, this approach clearly gives more accurate clustering estimates.
Nonetheless, large deviations from the reference values can still be seen, especially for higher order multipoles.    
It is also interesting to note that weighting the randoms is not
  effective on small scales $\lesssim 2 \ h^{-1}$Mpc, for the monopole and quadrupole (Fig. \ref{fig vamo}, green dashed). This might be, at least in part, a consequence of the fact that, on such scales, the number of galaxy pairs exceeds that of the random ones. Likely, the small-scale performance of this approach can be improved by using a denser random sample. For purely computational reasons, in this work we used a random sample about 5 times denser on average than the data. The transition scale at which the (weighted) data pairs exceed the random ones is about $7 \ h^{-1}$Mpc.

In principle, moving the clustering-independent fluctuations from the data to the randoms could have beneficial effects even when combined with PIP weights.
Specifically, we expect an improvement in the signal-to-noise ratio of the measurements, because the higher average density of the random sample means it has intrinsically lower shot noise than the galaxies. 
This topic is investigated in App. \ref{sec PIPvscomp}, where we show that the overall impact is negligible, at least for the DESI survey. 

In this work we do not consider more standard countermeasures to the missing observation problem, such as the widely used nearest neighbour (NN) upweighting (e.g. BOSS survey, \citealt{anderson2012}).
This technique consists of assigning the individual weight of a missing galaxy to its (angularly) closest (observed) companion. 
A specific comparison of the performance of PIP and NN is already provided in \citet{bianchi2017}.
As explained in that paper, the NN correction is formally included in the PIP description.
If, for example, we just focus on the problem of choosing a galaxy out of a pair with uniform probability, it is clear that the two approaches will give the same weight $w=2$ to the observed galaxy. 
Unfortunately, a generic targeting algorithm has to deal with more complex choices, which cannot be handled exactly by the simple NN prescription.

In Smith et al., currently under DESI internal review, the authors present an exhaustive test of different fibre-mitigation techniques, including NN upweighting, applied to mocks of the DESI Bright Galaxy Sample (section 3.1 of \citealt{amir2016a}). This survey will take place when moonlight prevents efficient observations of faint galaxies.
As expected, the NN correction performs reasonably well on BAO scales but, even when the survey is complete, it yields a systematic error larger than the intrinsic variance of the clustering on scales that can be relevant, e.g., for RSD analysis.

\subsection{Errors}\label{sec results: err}

One interesting question is the degradation of the error on the
clustering signal due to the fibre assignment. Fundamentally, the
issue is that, for some particular patterns of galaxies, we rely on a
small number of observations to tell us the clustering of a large
number of missed patterns. The PIP weights correctly upweight these
samples (for example increasing the importance of overlap regions for
small-separation pairs, to counterbalance their lack in other
regions), but they cannot correct for the fact that we fundamentally
lack signal. We can split the causes of the degradation into two,
although they are intricately linked: the lack of pairs on any
particular scale and orientation of the line-of-sight, and the
increase in the shot noise caused by the weighting scheme.

\begin{figure*}
 \begin{center}
   \includegraphics[width=5.8cm]{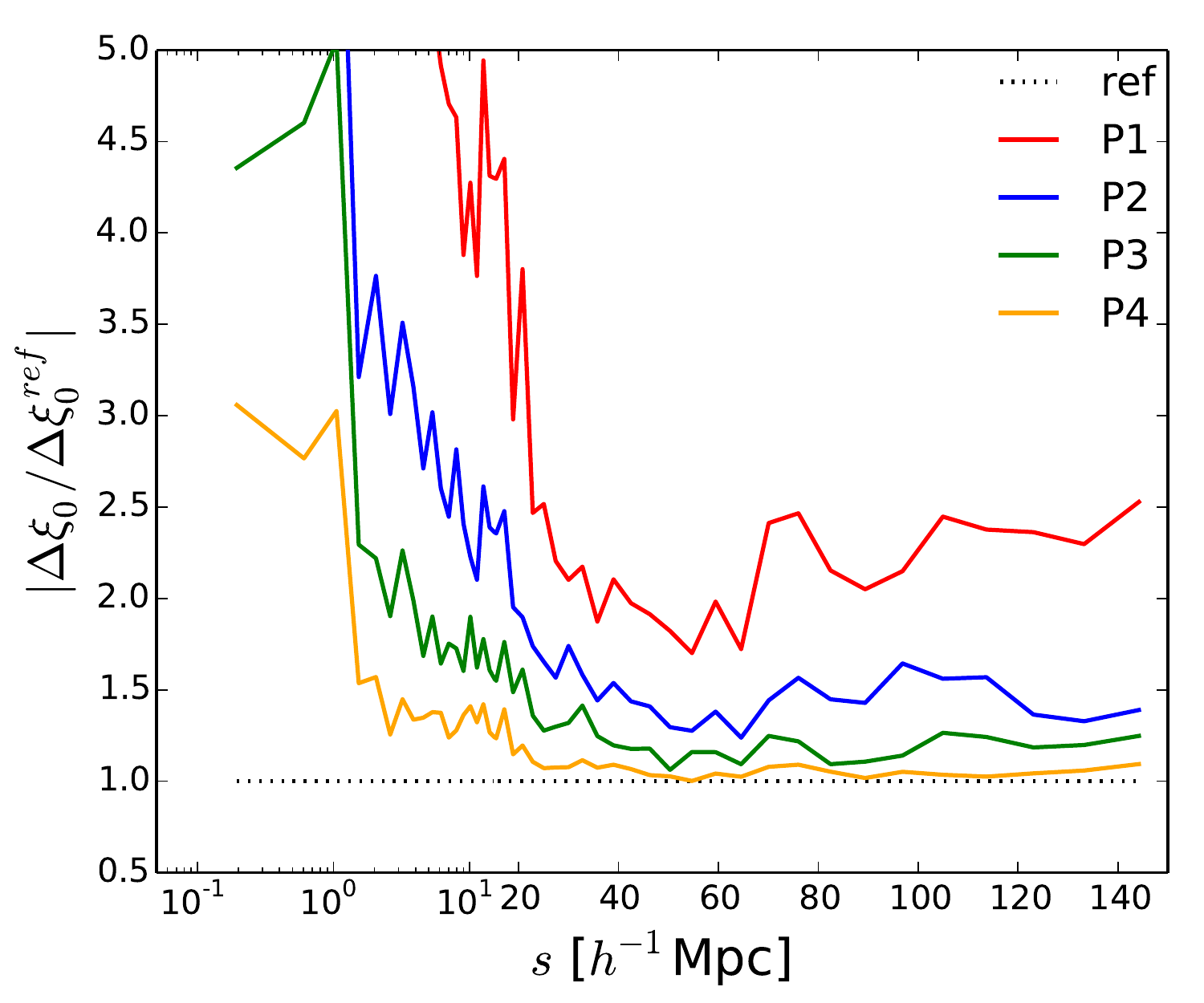}
   \includegraphics[width=5.8cm]{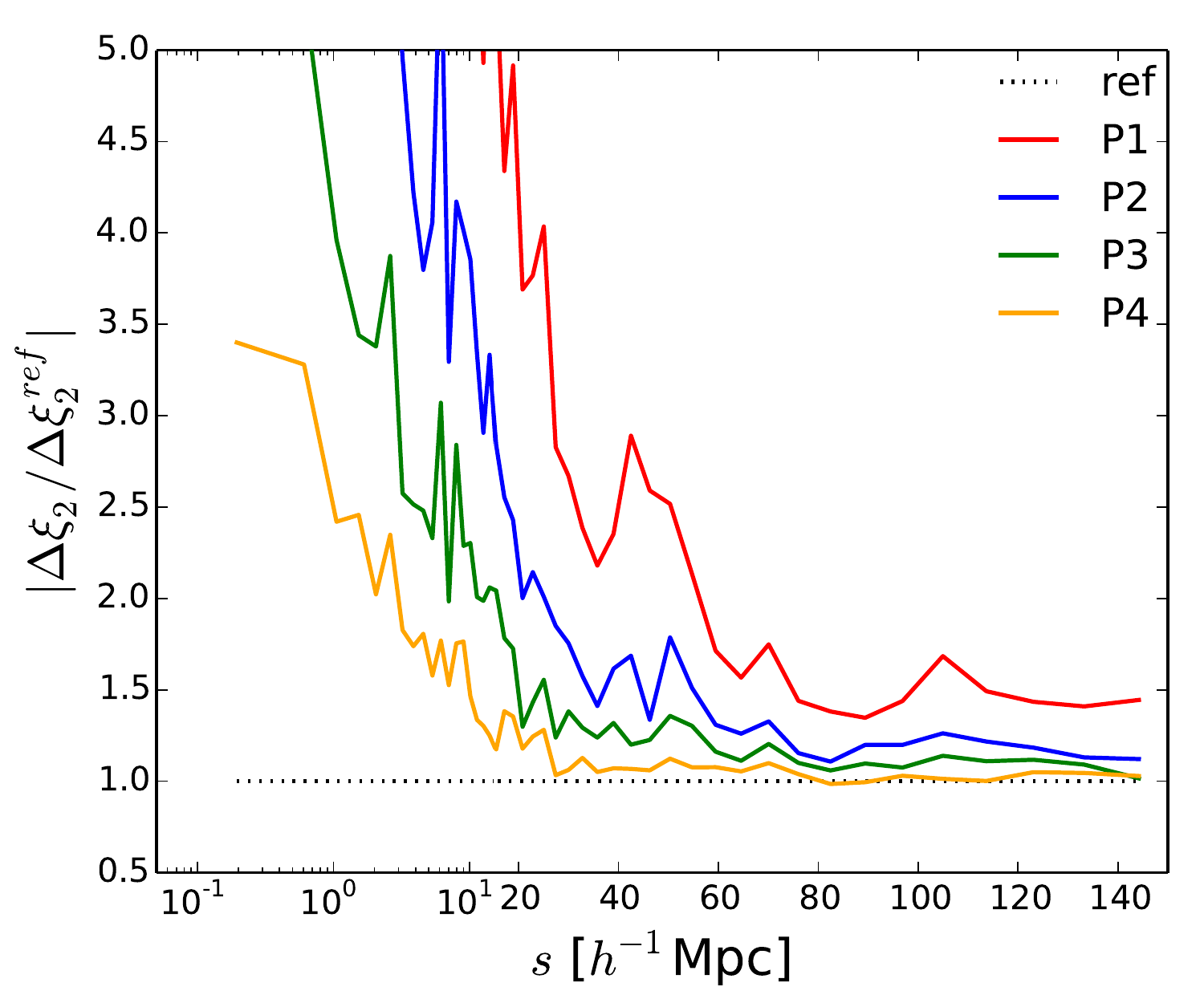}
   \includegraphics[width=5.8cm]{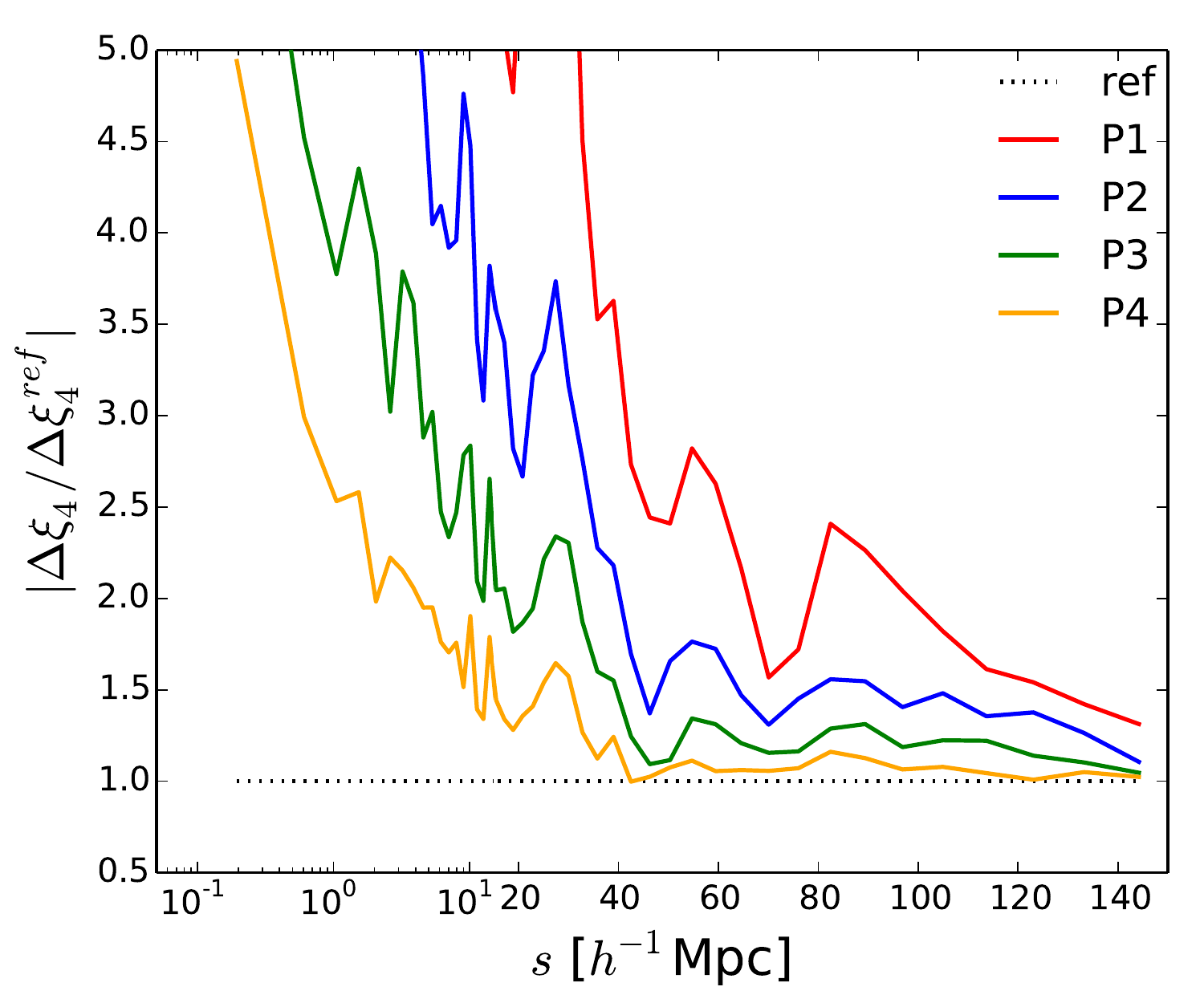}
   \caption{Error in the recovered multipole moments, compared to
     sample variance, calculated from the clustering of the parent
     sample.}
   \label{fig err1}
 \end{center}
\end{figure*}

The evolution of the error as a function of the number of passes of
the instrument is shown in Fig.~\ref{fig err1}, which reports the
standard deviation of the measured multipoles $\Delta \xi_n$ divided
by that of a reference value corresponding to the intrinsic scatter in the mocks without fibre assignment.
On large-scales we see that the measurement error converges to that of the parent sample, while on
small scales, the observing effects leave a significant hit from the
fibre assignment, even after four passes.
This test can be seen an ``unfair'' comparison with an unrealistic survey in which we are able to target all available galaxies.
As such, it just provides us with a lower bound.
From this perspective, it is reassuring to see how, even in this ``unfair'' comparison, the error hits the lower bound, essentially becoming cosmic-variance dominated, when scales and number of passes grow.  

\begin{figure*}
 \begin{center}
   \includegraphics[width=5.8cm]{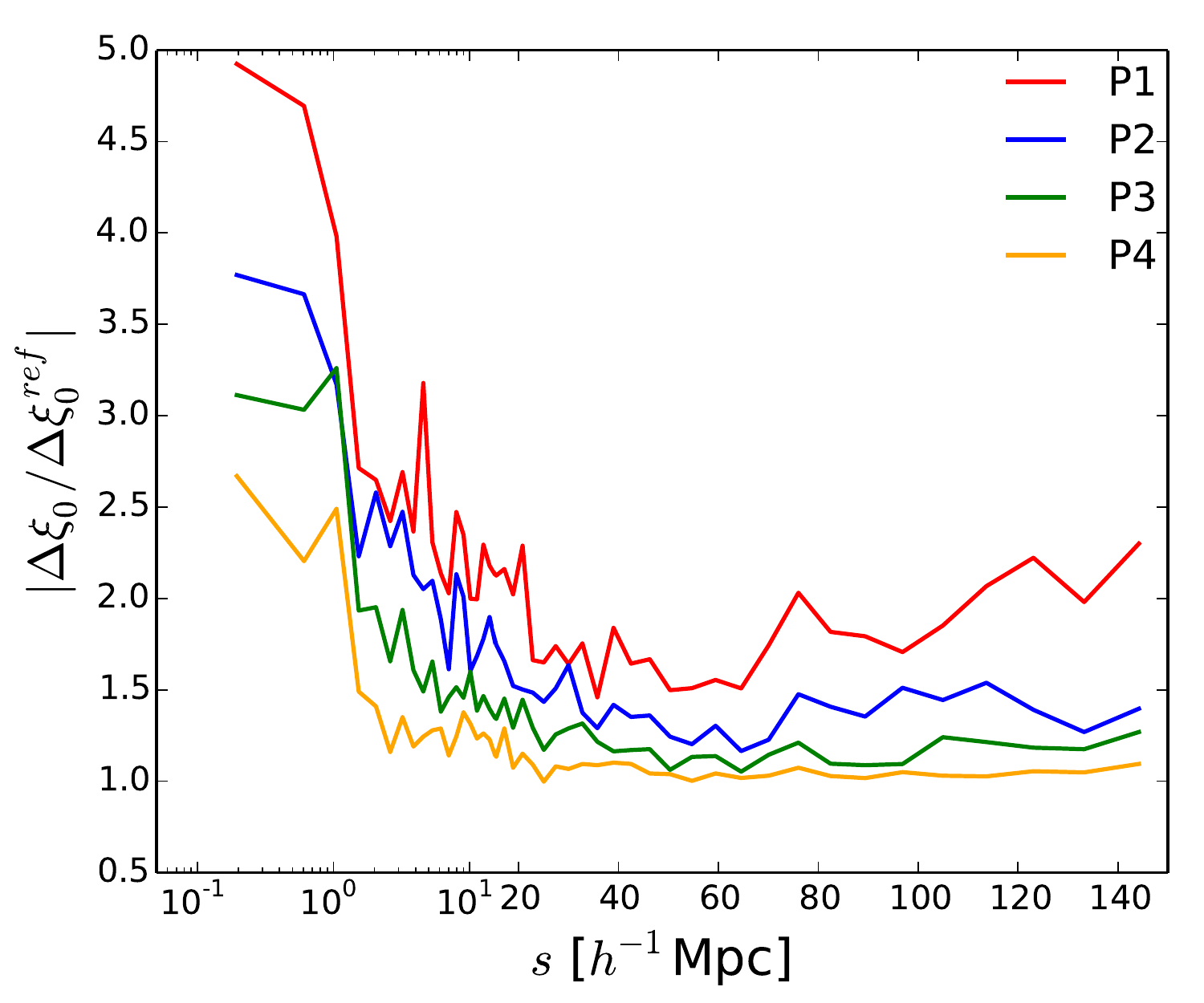}
   \includegraphics[width=5.8cm]{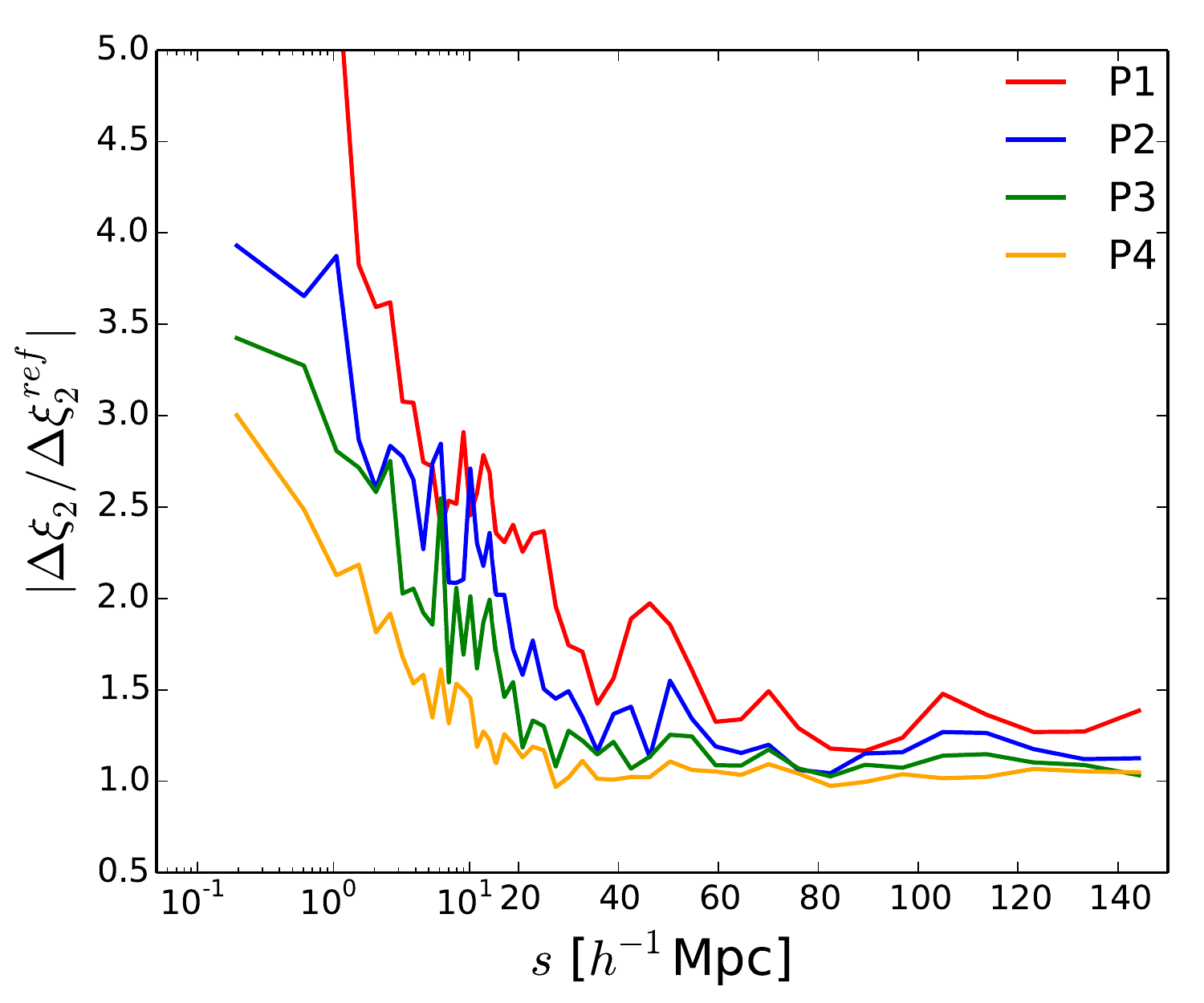}
   \includegraphics[width=5.8cm]{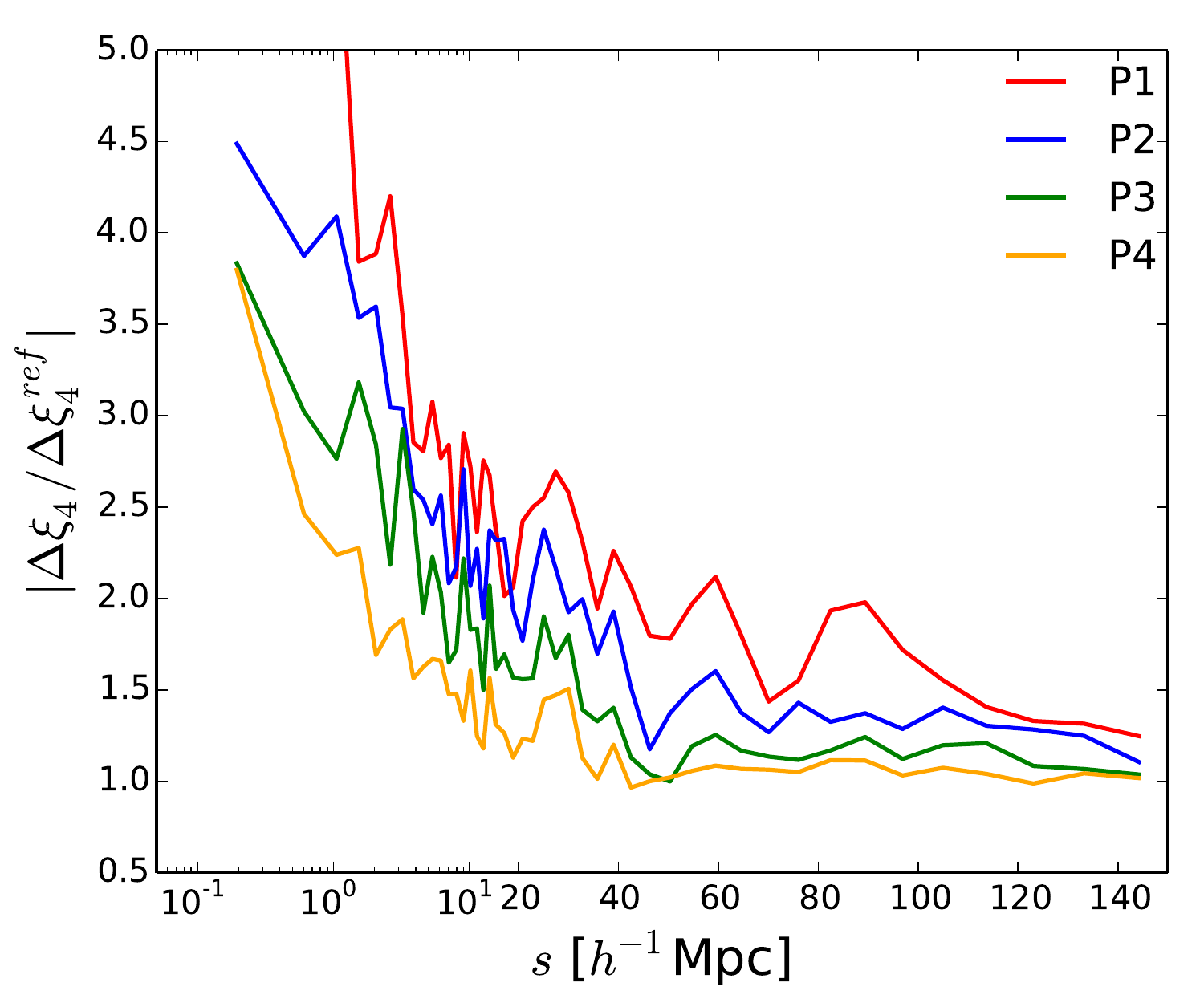}
   \caption{Error in the recovered multipole moments, compared to that recovered from a set of catalogues obtained by uniformly diluting the mocks until they match the number density of their fibre-assigned counterparts.}
   \label{fig err2}
 \end{center}
\end{figure*}

Rather than comparing to a single reference set of correlation
functions, in Fig.~\ref{fig err2}, we compare the error from our
recovered PIP-corrected correlation functions to that from samples
where, for each set of passes we randomly select the same number of
galaxies from the target population. Thus, by comparing the two, we
isolate the degradation from the fibre assignment from that of the
changing overall galaxy density. For a survey created from one pass of
the instrument, we see a factor of $\sim$2 increase in the error from
the fibre assignment algorithm, and this reduces as the number of
passes increases.
 
Despite the encouraging results of the previous tests, it would still be worthwhile to see if the error can be reduced further.
As anticipated above, we already tested probably the most obvious extension, which consists of taking advantage of the higher density of the random sample to reduce the noise, obtaining no appreciable improvement (App. \ref{sec PIPvscomp}).
On the other hand, it is certainly possible to imagine very idealised scenarios in which the bias on the clustering caused by missing observations can be modelled theoretically, in a deterministic way.
In such a case we could get unbiased estimates of the cosmological parameters just by fitting the unweighted correlation function, which, by construction, has smaller error bars.
This would result in smaller errors on the cosmological parameters, at least if the model does not require nuissance parameters and does not discard information. 
It is nonetheless doubtful that such an approach can be self consistently applied to a realistic algorithm/survey and, even more, to all possible algorithms/surveys.        
We leave this general discussion about what is an optimal estimator in case of missing observation for further work.

\subsection{BAO peak}\label{sec results: BAO}

The obvious further step is to assess how well the PIP scheme performs in recovering the position of the BAO peak.
To this purpose, we follow \citet{anderson2014a, anderson2014b} and fit the correlation function to
\begin{equation}
\xi_m(r) = B^2 \xi_{ref}(r\alpha) + \frac{a_1}{r^2} +\frac{a_2}{r} +a_3 \ ,
\end{equation}
where $B$ and $a_i$ are nuisance parameters that adjust the amplitude and the broadband shape of the correlation function, respectively.
The parameter $\alpha$ encodes any shifts in the BAO feature compared to the reference model $\xi_{ref}$, which is what we want to assess with this test.

Since we only have 25 mock catalogues, when we fit for $\alpha$, instead of using the covariance matrix estimated from the mocks we adopt an effective covariance matrix, obtained by rescaling the publicly available covariance matrix of the DR11 CMASS BOSS galaxy sample1: $C_{eff} = V_{BOSS} / V \ C_{BOSS}$.
This is essentially the same approach adopted by \citet{burden2017} but here the purpose of the test is different.
In that work the authors measure $\alpha$ from the full mocks, i.e. without fibre assignment, as a way to show that the model they propose is unbiased.
Since if there is no fibre assignment the PIP correction vanishes, i.e. trivially each pair has unitary weight, the recovered clustering is unbiased by construction.
Therefore, in this work we test how well we can recover $\alpha$ in the actual presence of fibre assignment.

Unfortunately, the covariance matrix that we are using cannot be considered a realistic alternative to the ``true'' one, for the following reasons.
As a zero-order effect, even after volume rescaling, the average number density depends on the number of passes of the instruments, and, especially after one or two passes it is significantly lower than that of the BOSS survey.
Furthermore, even if we managed to correct this deficit, we would still have higher order effects coming from the fact that we are not using the DESI fibre-assignment algorithm coupled to the PIP weights to derive the covariance matrix, but, instead, their BOSS counterparts.
Similarly to the average-number-density issue, this inconsistency is likely to become more important as the number of passes decreases, i.e. when the imprint of the DESI instrument becomes dominant.
As a way to keep these limitations under control, we also process under the same pipeline the uniformly diluted samples already adopted as a lower bound for the error on the multipoles in Sec. \ref{sec results: err}.

\begin{figure*}
 \begin{center}
   \includegraphics[width=4.3cm]{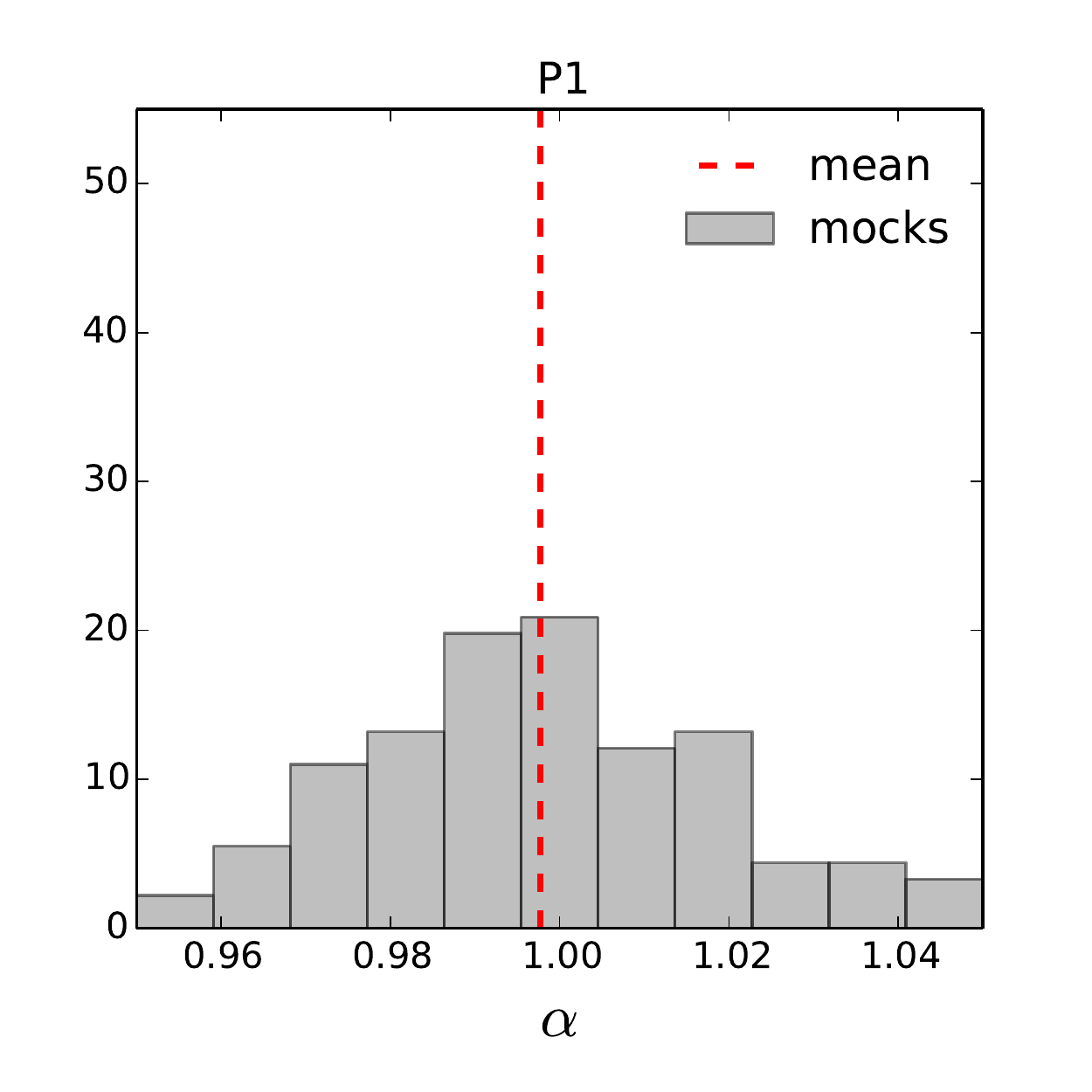}
   \includegraphics[width=4.3cm]{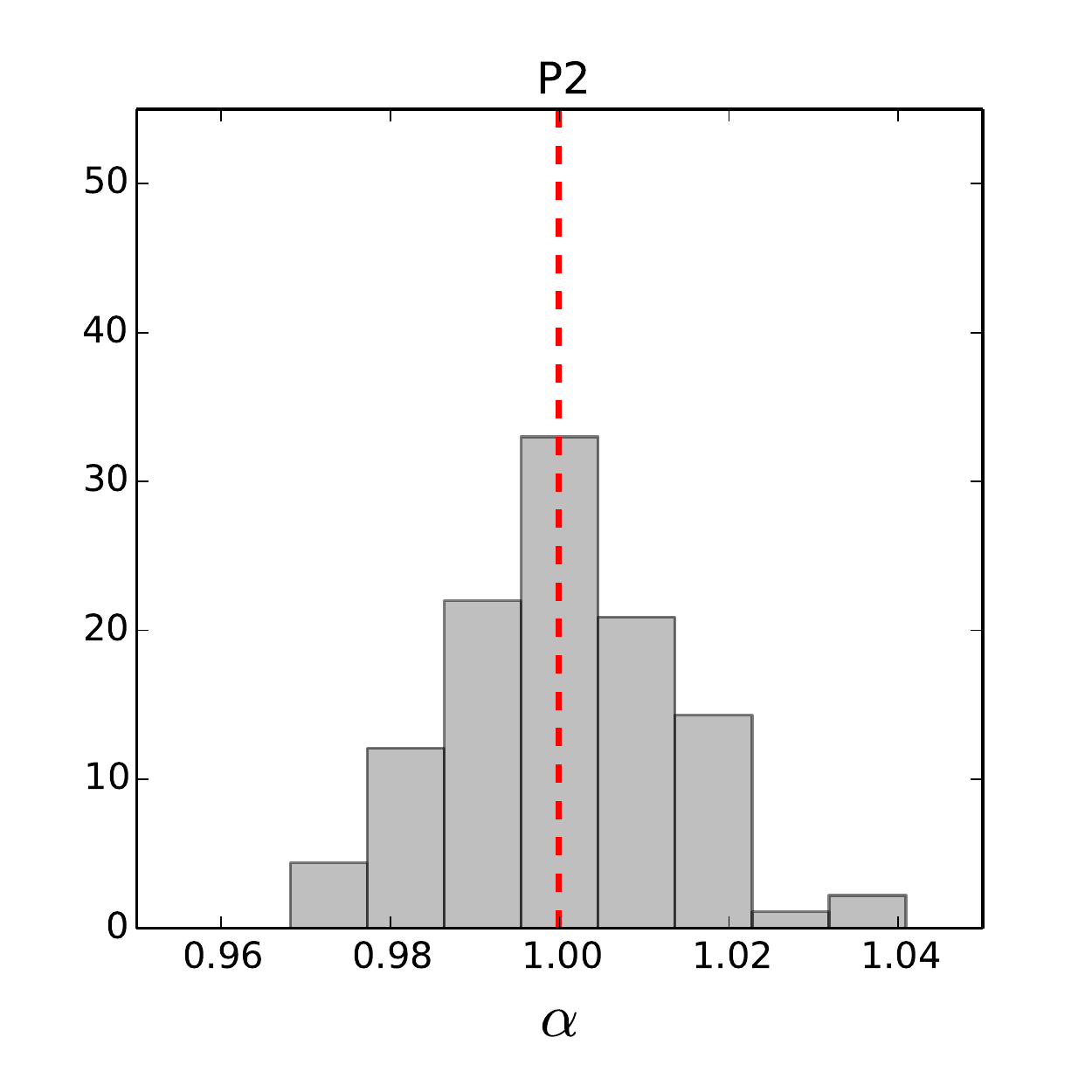}   
   \includegraphics[width=4.3cm]{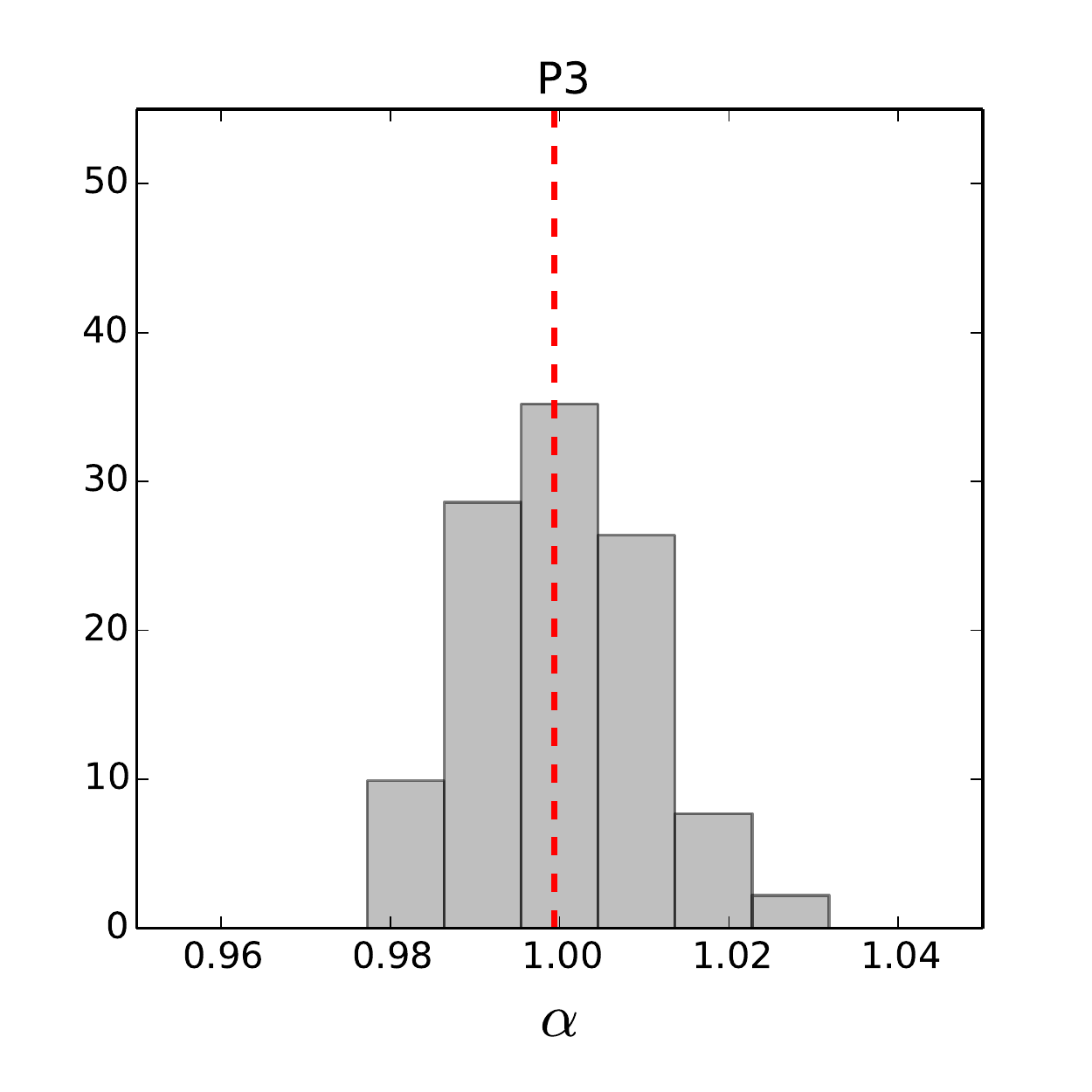}
   \includegraphics[width=4.3cm]{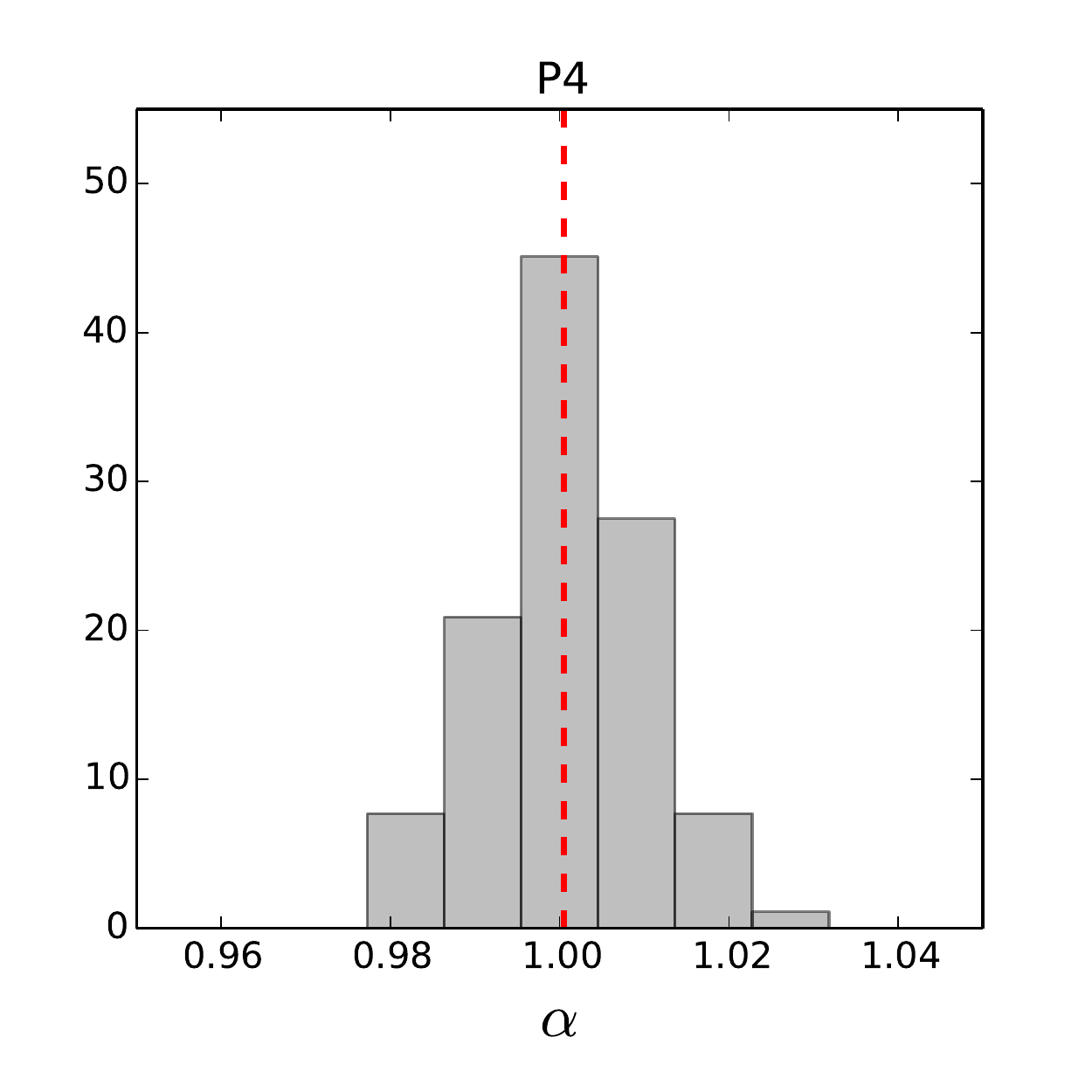} \\  
   \includegraphics[width=4.3cm]{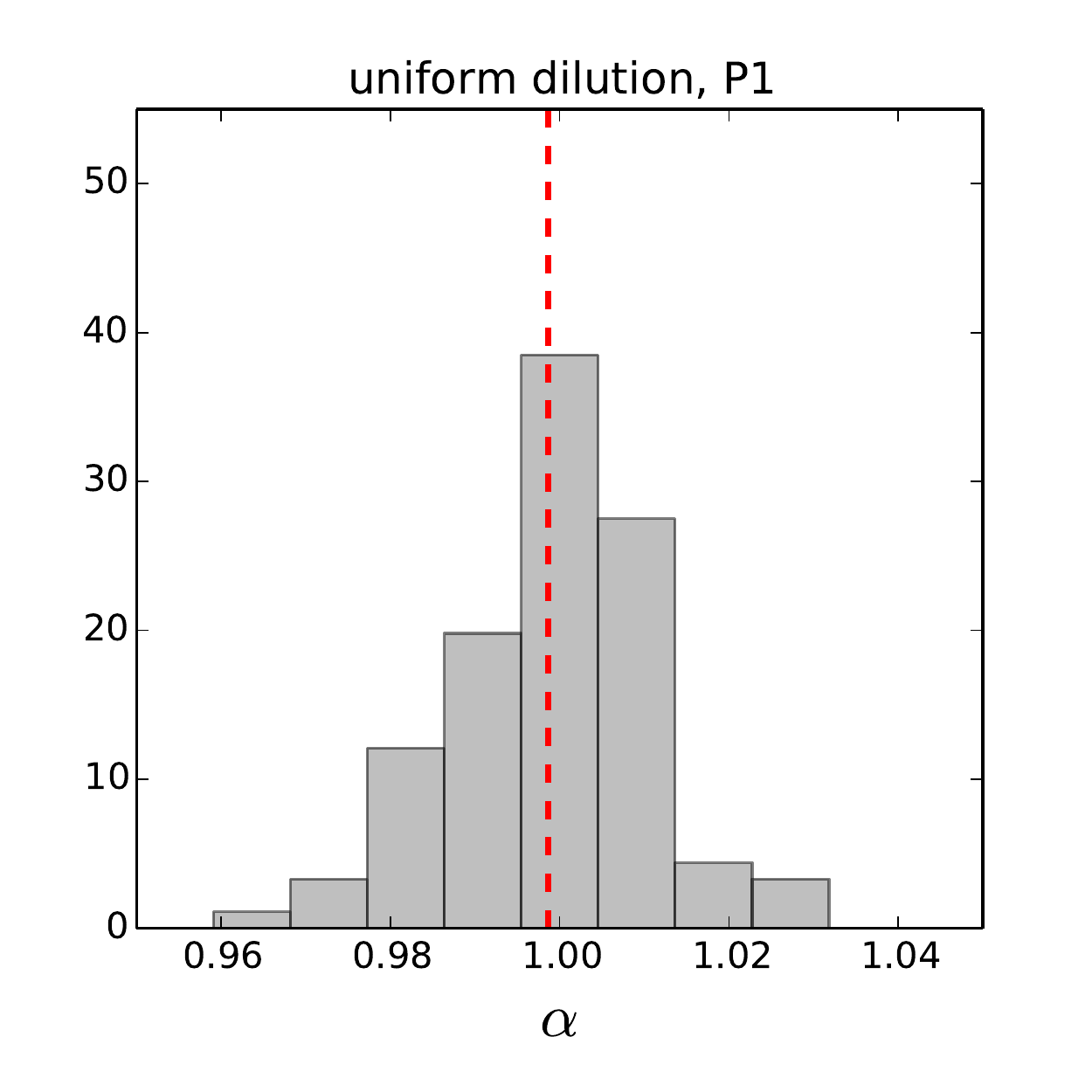}
   \includegraphics[width=4.3cm]{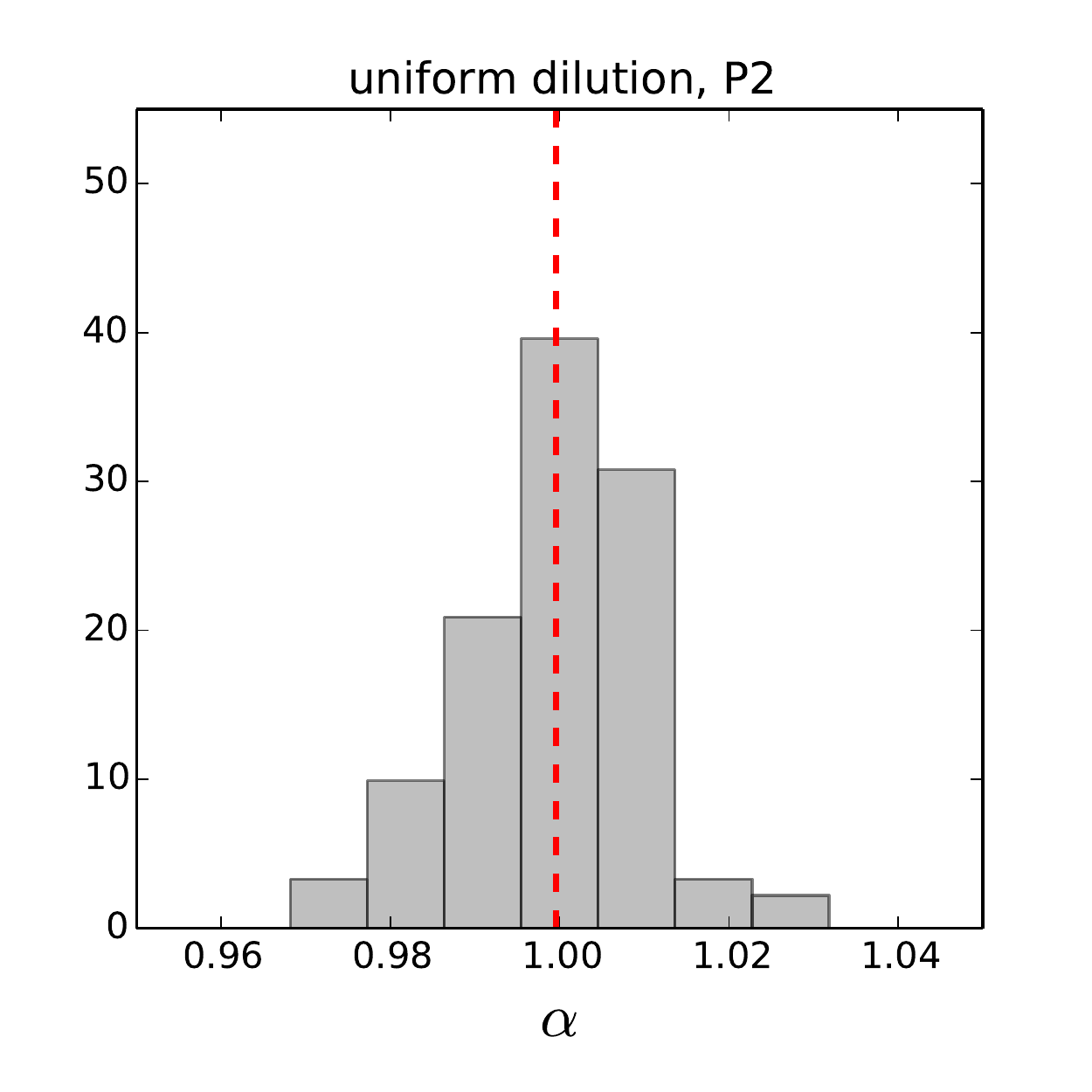}   
   \includegraphics[width=4.3cm]{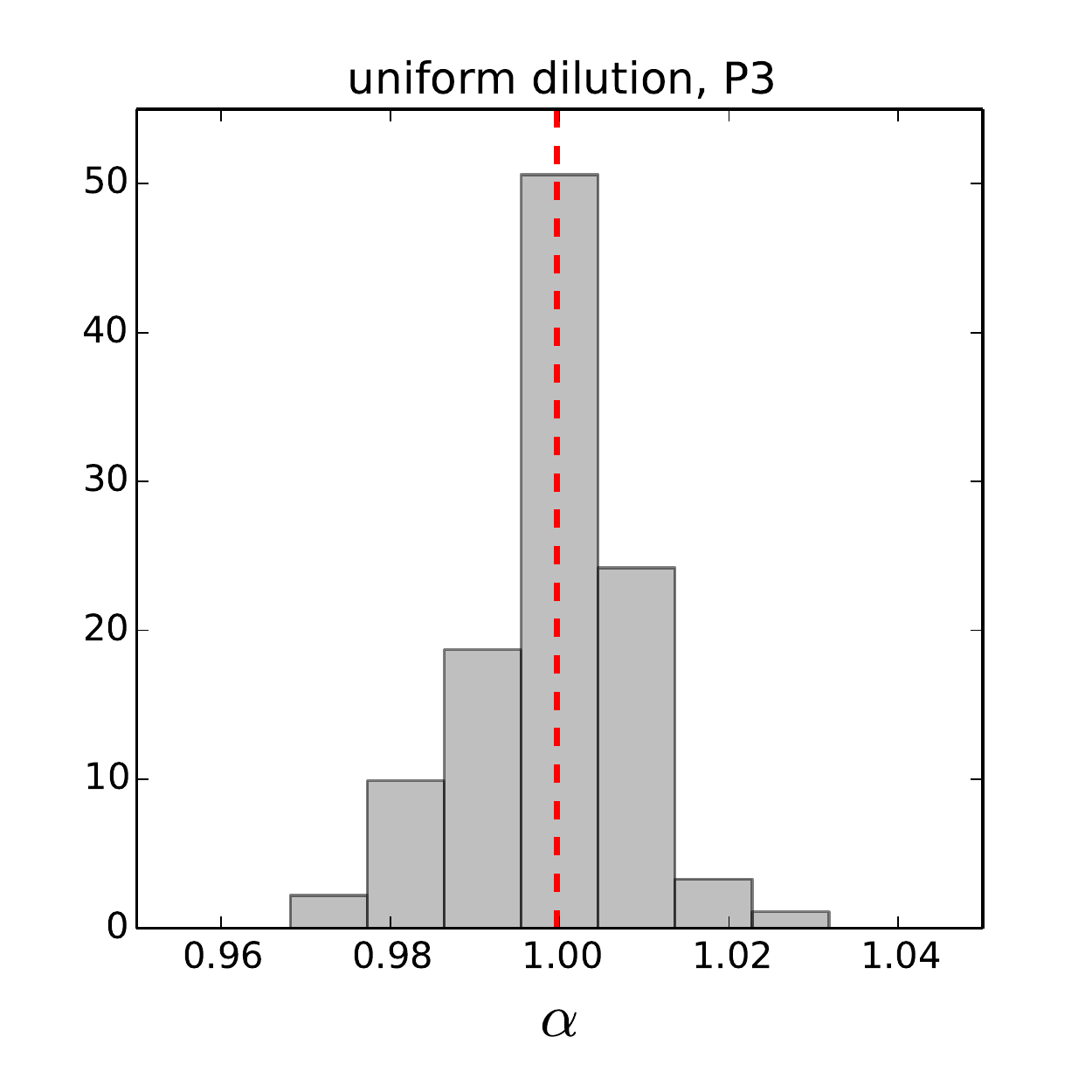}
   \includegraphics[width=4.3cm]{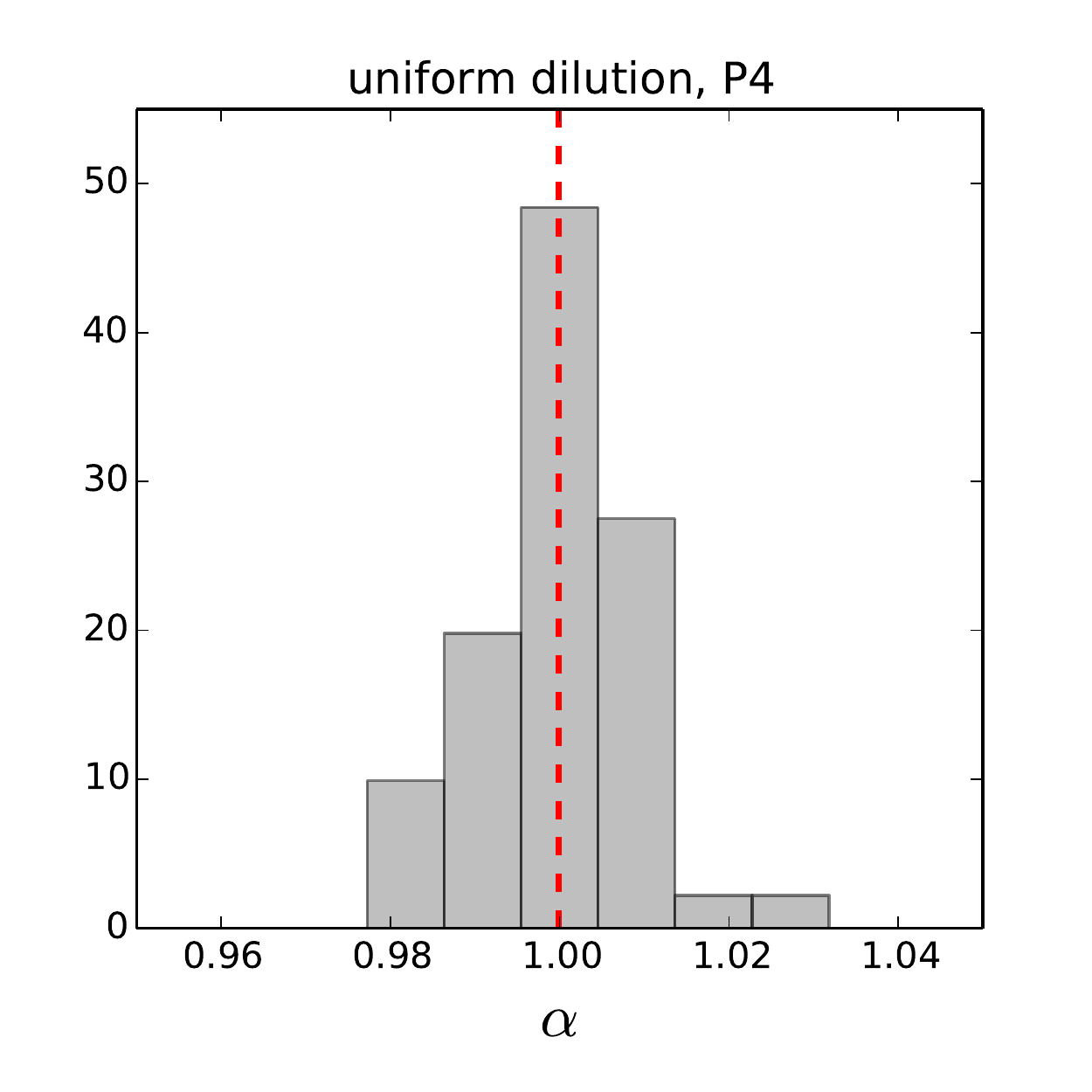}   
   \caption{Upper row: distribution of the best-fit $\alpha$ across different mocks for different number of passes of the instrument, as labelled in the figure. For comparison, in the lower row we report the same measurements in the idealised case of uniform dilution.}
   \label{fig alpha}
 \end{center}
\end{figure*}

In the upper row of Fig. \ref{fig alpha} we show histograms of the best-fit $\alpha$ from all the $25 \times 4 = 100$ realisations discussed in Sec. \ref{sec results: xi}, for the first four passes of the instrument.
All the measurements are clearly unbiased.
As for the error, due to the complications mentioned above, we limit our discussion to the comparison between DESI-targeting-plus-PIP-weights and uniform-dilution case (lower row).
The relative behaviour resembles that already observed for the multipoles, with the ratio of the standard deviations $R=\sigma_{PIP} / \sigma_{unif}$ being $\{R_{P1}, R_{P2}, R_{P3}, R_{P4}\} \approx \{1.76, 1.27, 1.09, 1.06\}$.

\section{Conclusions}\label{sec conclusions}

We have applied the algorithm to correct for missing observations in
galaxy surveys, developed in \citet{bianchi2017}, to mock catalogues
designed to mimic the Emission Line Galaxy sample to be obtained from
the upcoming DESI experiment. We find that the algorithm correctly
deals with the effects of fibre assignment, resulting in unbiased
correlation function measurements. The weighted correlation function
exhibits worse noise compared to a random sampling of
galaxies. However we find that this degradation has effectively
vanished after three passes of the instrument over the sky, and similarly for the position of the BAO peak.

The work presented here has proven the utility of the algorithm for
correcting for fibre assignment within DESI, and paves the way for
survey design knowing that we can correct for these
effects. There is an increase in the expected errors, compared to a
random sampling of galaxies, but this decreases with number of
passes. We should expect that, by the end of the survey, there is
no significant degradation on the BAO scales, so errors should match
those expected.

Two questions remain for the full application of the PIP-weighting:
how reconstruction of the linear density field affects the correction,
and how to apply the algorithm in Fourier-space. These are left for
future work.

\section*{Acknowledgements}

We thank Martin White for providing the mock catalogues and Shaun Cole for useful discussions.
DB and WJP acknowledge support from the European Research Council
through the Darksurvey grant 614030.  WJP also acknowledges support
from the UK Science and Technology Facilities Council grant
ST/N000668/1 and the UK Space Agency grant ST/N00180X/1.
JEFR acknowledges support from COLCIENCIAS Contract No. 287-2016, Project 1204-712-50459.
This research is supported by the Director, Office of Science, Office of High Energy Physics of the U.S. Department of Energy under Contract No. 
DE-AC02-05CH1123, and by the National Energy Research Scientific Computing Center, a DOE Office of Science User Facility under the same 
contract; additional support for DESI is provided by the U.S. National Science Foundation, Division of Astronomical Sciences under Contract No. 
AST-0950945 to the National Optical Astronomy Observatory; the Science and Technologies Facilities Council of the United Kingdom; the Gordon 
and Betty Moore Foundation; the Heising-Simons Foundation; the National Council of Science and Technology of Mexico, and by the DESI 
Member Institutions.  The authors are honored to be permitted to conduct astronomical research on Iolkam Du'ag (Kitt Peak), a mountain with 
particular significance to the Tohono O'odham Nation. 




\bibliographystyle{mnras}
\bibliography{./biblio_db}

\begin{thebibliography}{}
\makeatletter
\relax
\def\mn@urlcharsother{\let\do\@makeother \do\$\do\&\do\#\do\^\do\_\do\%\do\~}
\def\mn@doi{\begingroup\mn@urlcharsother \@ifnextchar [ {\mn@doi@}
  {\mn@doi@[]}}
\def\mn@doi@[#1]#2{\def\@tempa{#1}\ifx\@tempa\@empty \href
  {http://dx.doi.org/#2} {doi:#2}\else \href {http://dx.doi.org/#2} {#1}\fi
  \endgroup}
\def\mn@eprint#1#2{\mn@eprint@#1:#2::\@nil}
\def\mn@eprint@arXiv#1{\href {http://arxiv.org/abs/#1} {{\tt arXiv:#1}}}
\def\mn@eprint@dblp#1{\href {http://dblp.uni-trier.de/rec/bibtex/#1.xml}
  {dblp:#1}}
\def\mn@eprint@#1:#2:#3:#4\@nil{\def\@tempa {#1}\def\@tempb {#2}\def\@tempc
  {#3}\ifx \@tempc \@empty \let \@tempc \@tempb \let \@tempb \@tempa \fi \ifx
  \@tempb \@empty \def\@tempb {arXiv}\fi \@ifundefined
  {mn@eprint@\@tempb}{\@tempb:\@tempc}{\expandafter \expandafter \csname
  mn@eprint@\@tempb\endcsname \expandafter{\@tempc}}}

\bibitem[\protect\citeauthoryear{{Anderson} et~al.,}{{Anderson}
  et~al.}{2012}]{anderson2012}
{Anderson} L.,  et~al., 2012, \mn@doi [\mnras]
  {10.1111/j.1365-2966.2012.22066.x}, \href
  {http://adsabs.harvard.edu/abs/2012MNRAS.427.3435A} {427, 3435}

\bibitem[\protect\citeauthoryear{{Anderson} et~al.,}{{Anderson}
  et~al.}{2014a}]{anderson2014a}
{Anderson} L.,  et~al., 2014a, \mn@doi [\mnras] {10.1093/mnras/stt2206}, \href
  {http://adsabs.harvard.edu/abs/2014MNRAS.439...83A} {439, 83}

\bibitem[\protect\citeauthoryear{{Anderson} et~al.,}{{Anderson}
  et~al.}{2014b}]{anderson2014b}
{Anderson} L.,  et~al., 2014b, \mn@doi [\mnras] {10.1093/mnras/stu523}, \href
  {http://adsabs.harvard.edu/abs/2014MNRAS.441...24A} {441, 24}

\bibitem[\protect\citeauthoryear{{Bianchi} \& {Percival}}{{Bianchi} \&
  {Percival}}{2017}]{bianchi2017}
{Bianchi} D.,  {Percival} W.~J.,  2017, \mn@doi [\mnras]
  {10.1093/mnras/stx2053}, \href
  {http://adsabs.harvard.edu/abs/2017MNRAS.472.1106B} {472, 1106}

\bibitem[\protect\citeauthoryear{{Burden}, {Padmanabhan}, {Cahn}, {White}  \&
  {Samushia}}{{Burden} et~al.}{2017}]{burden2017}
{Burden} A.,  {Padmanabhan} N.,  {Cahn} R.~N.,  {White} M.~J.,   {Samushia} L.,
   2017, \mn@doi [\jcap] {10.1088/1475-7516/2017/03/001}, \href
  {http://adsabs.harvard.edu/abs/2017JCAP...03..001B} {3, 001}

\bibitem[\protect\citeauthoryear{{Cahn}, {Bailey}, {Dawson}, {Forero Romero},
  {Schlegel}, {White}  \& {DESI}}{{Cahn} et~al.}{2015}]{cahn2015}
{Cahn} R.~N.,  {Bailey} S.~J.,  {Dawson} K.~S.,  {Forero Romero} J.,
  {Schlegel} D.~J.,  {White} M.,   {DESI} 2015, in American Astronomical
  Society Meeting Abstracts. p. 336.10

\bibitem[\protect\citeauthoryear{{DESI Collaboration} et~al.,}{{DESI
  Collaboration} et~al.}{2016a}]{amir2016a}
{DESI Collaboration} et~al., 2016a, preprint, \href
  {http://adsabs.harvard.edu/abs/2016arXiv161100036D} {} (\mn@eprint {arXiv}
  {1611.00036})

\bibitem[\protect\citeauthoryear{{DESI Collaboration} et~al.,}{{DESI
  Collaboration} et~al.}{2016b}]{amir2016b}
{DESI Collaboration} et~al., 2016b, preprint, \href
  {http://adsabs.harvard.edu/abs/2016arXiv161100037D} {} (\mn@eprint {arXiv}
  {1611.00037})

\bibitem[\protect\citeauthoryear{{Dawson} et~al.,}{{Dawson}
  et~al.}{2013}]{dawson2013}
{Dawson} K.~S.,  et~al., 2013, \mn@doi [\aj] {10.1088/0004-6256/145/1/10},
  \href {http://adsabs.harvard.edu/abs/2013AJ....145...10D} {145, 10}

\bibitem[\protect\citeauthoryear{{Gonzalez-Perez} et~al.,}{{Gonzalez-Perez}
  et~al.}{2018}]{Gonzalez-perez2017}
{Gonzalez-Perez} V.,  et~al., 2018, \mn@doi [\mnras] {10.1093/mnras/stx2807},
  \href {http://adsabs.harvard.edu/abs/2018MNRAS.474.4024G} {474, 4024}

\bibitem[\protect\citeauthoryear{{Landy} \& {Szalay}}{{Landy} \&
  {Szalay}}{1993}]{landy1993}
{Landy} S.~D.,  {Szalay} A.~S.,  1993, \mn@doi [\apj] {10.1086/172900}, \href
  {http://adsabs.harvard.edu/abs/1993ApJ...412...64L} {412, 64}

\bibitem[\protect\citeauthoryear{{Masjedi} et~al.,}{{Masjedi}
  et~al.}{2006}]{masjedi2006}
{Masjedi} M.,  et~al., 2006, \mn@doi [\apj] {10.1086/503536}, \href
  {http://adsabs.harvard.edu/abs/2006ApJ...644...54M} {644, 54}

\bibitem[\protect\citeauthoryear{{Percival} \& {Bianchi}}{{Percival} \&
  {Bianchi}}{2017}]{percival2017}
{Percival} W.~J.,  {Bianchi} D.,  2017, \mn@doi [\mnras]
  {10.1093/mnrasl/slx135}, \href
  {http://adsabs.harvard.edu/abs/2017MNRAS.472L..40P} {472, L40}

\bibitem[\protect\citeauthoryear{{Pinol}, {Cahn}, {Hand}, {Seljak}  \&
  {White}}{{Pinol} et~al.}{2017}]{pinol2017}
{Pinol} L.,  {Cahn} R.~N.,  {Hand} N.,  {Seljak} U.,   {White} M.,  2017,
  \mn@doi [\jcap] {10.1088/1475-7516/2017/04/008}, \href
  {http://adsabs.harvard.edu/abs/2017JCAP...04..008P} {4, 008}

\bibitem[\protect\citeauthoryear{{Tinker} et~al.,}{{Tinker}
  et~al.}{2012}]{tinker2012}
{Tinker} J.~L.,  et~al., 2012, \mn@doi [\apj] {10.1088/0004-637X/745/1/16},
  \href {http://adsabs.harvard.edu/abs/2012ApJ...745...16T} {745, 16}

\bibitem[\protect\citeauthoryear{{White}, {Tinker}  \& {McBride}}{{White}
  et~al.}{2014}]{white2014}
{White} M.,  {Tinker} J.~L.,   {McBride} C.~K.,  2014, \mn@doi [\mnras]
  {10.1093/mnras/stt2071}, \href
  {http://adsabs.harvard.edu/abs/2014MNRAS.437.2594W} {437, 2594}

\makeatother
\end{thebibliography}




\appendix

\section{Details on the measurements}\label{sec details}

The multipoles are obtained by first measuring $\xi(s, \mu)$ and then projecting it on the Legendre polynomials $\xi_l(s) = (2l+1)\int d\mu \ \xi(s, \mu) L_l(\mu)$.
For $\mu$ we split the interval $[0,1]$ into $100$ linear bins.
For $s$ we use the shifted logarithmic binning scheme introduced in \citet{bianchi2017},
\begin{equation}
s_i = 10^{x_0 + (i-1) \Delta x} - s_{sh} \ .
\end{equation}
Specifically we adopt $x_0 \approx 0.72$, $\Delta x \approx 0.033$ and
shift factor $s_{sh} = 5\mpcoh$.  For the angular pair counts $DD_a(\theta)$ and $DR_a(\theta)$ we
use the same binning scheme with $x_0 \approx -0.96$,
$\Delta x \approx 0.085$ and $s_{sh} = 0.1$deg.

\section{Completeness map}\label{sec comp map}

The clustering-independent completeness map is obtained as follows.
\begin{enumerate}
\item\label{item maps}
We create an angular density map for each of the 25 mocks, without fibre assignment.
\item\label{item avg map}
We average over the so obtained 25 maps.
\item\label{item fa maps}
We create an angular density map for four different realisations of the fibre-assignment algorithm, for each of the 25 mocks.
\item\label{item avg fa map}
We average over the so obtained $100 =4 \times 25$ maps.
\item\label{item ratio}
We take the ratio between \ref{item avg fa map} and \ref{item avg map}.
\end{enumerate}
We repeat \ref{item fa maps}, \ref{item avg fa map} and \ref{item ratio} for four passes of the instrument.
For the density maps we use an adaptive grid to avoid the complication of having grid points with no galaxies.
This can be seen as a smoothing of the field in which the width of the filter is allowed to grow when the density drops below a given treshold.
One obvious alternative is to run the random sample trough the fibre-assignment algorithm, see e.g. \citet{burden2017}.
The advantage of the procedure adopted here is that the entanglement between clustering and targeting is more properly taken into account.
The main limitation is that, formally, in order to completely remove the clustering signal, we need an infinite number of mocks.

\begin{figure*}
 \begin{center}
   \includegraphics[width=8.0cm]{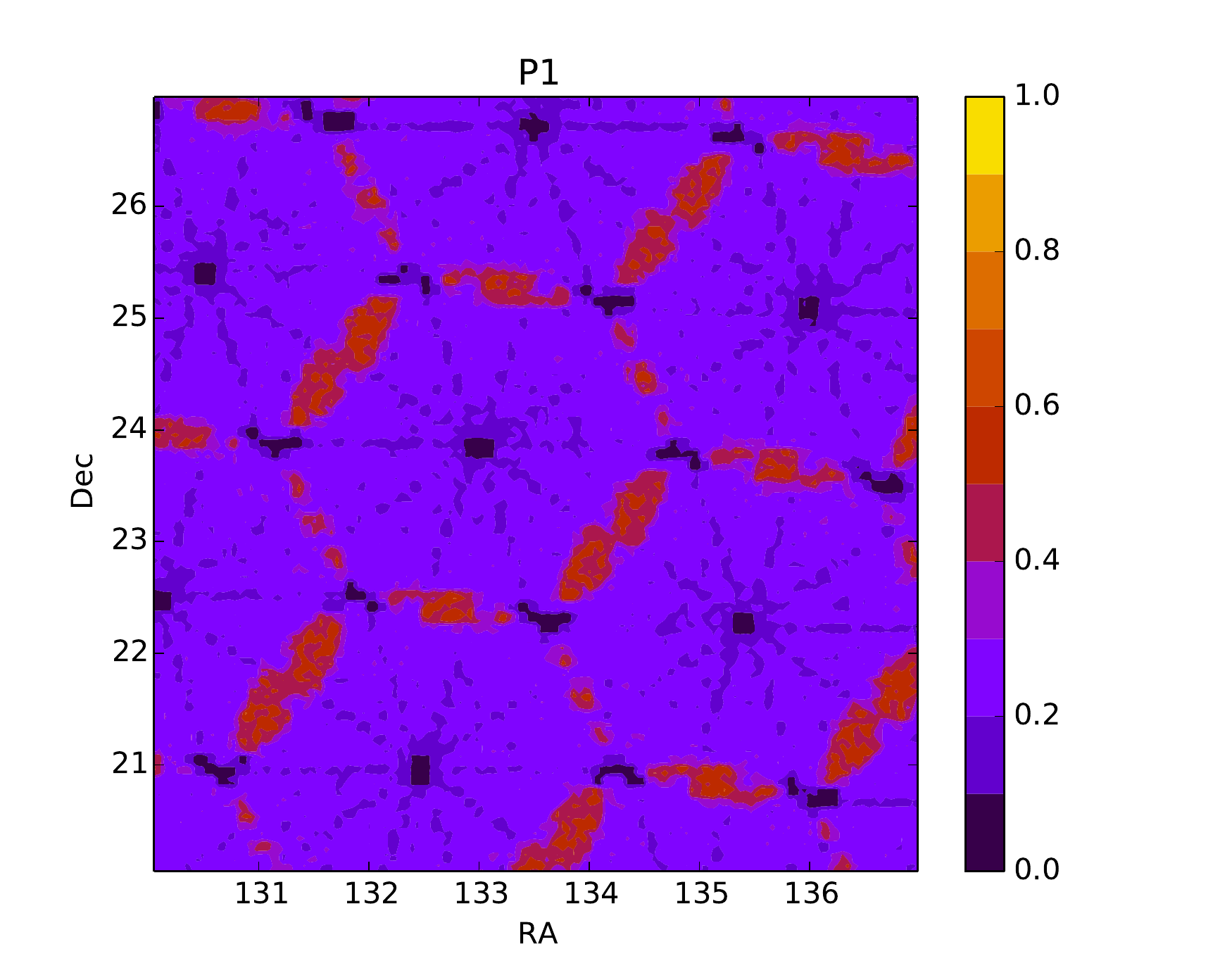}
   \includegraphics[width=8.0cm]{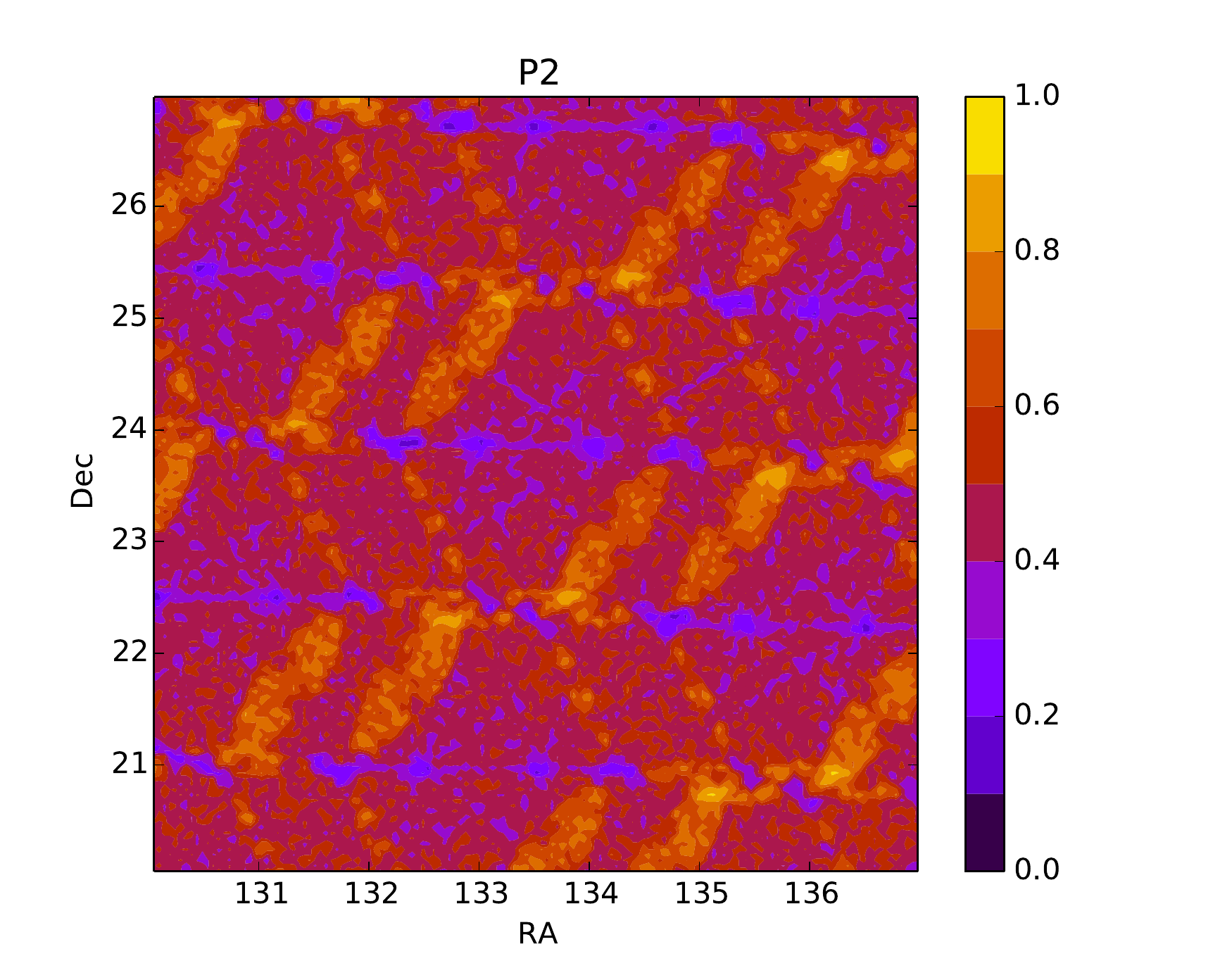}\\
   \includegraphics[width=8.0cm]{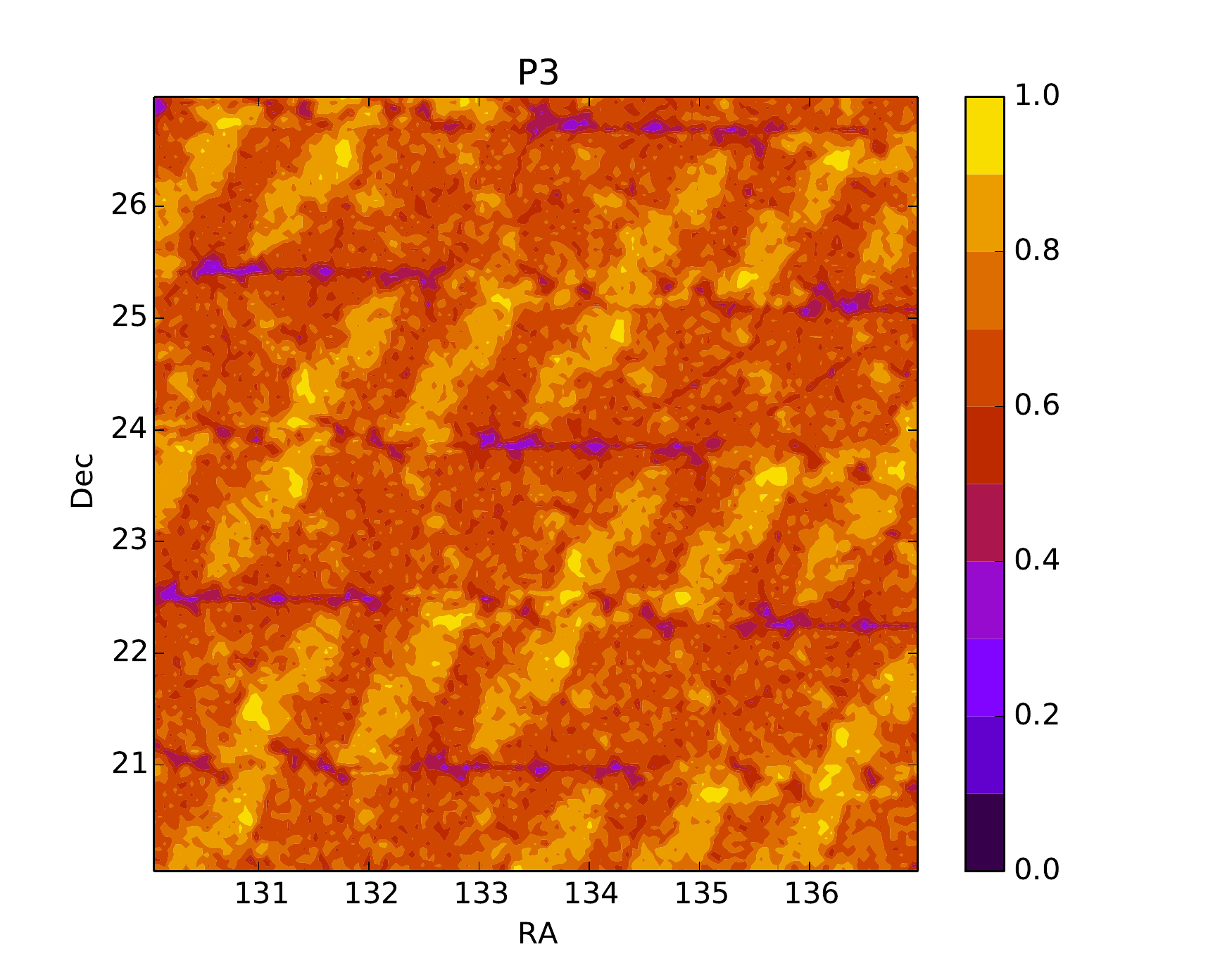}
   \includegraphics[width=8.0cm]{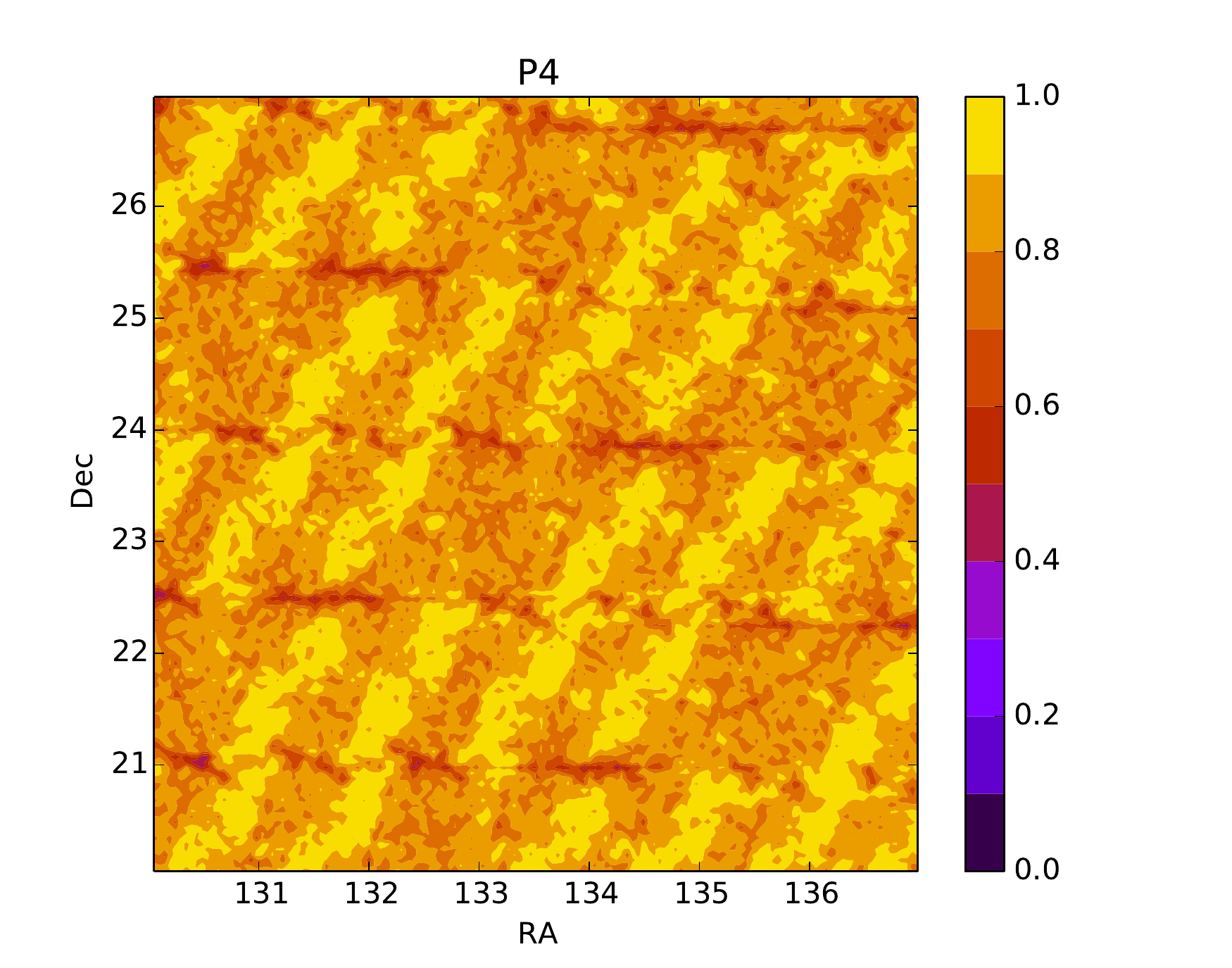}
   \caption{Clustering-independent completeness maps for different passes of the DESI instrument.}
   \label{fig comp map}
 \end{center}
\end{figure*}

The completeness maps are shown in Fig. \ref{fig comp map}.
The imprint of the the DESI instrument and its evolution with the number of passes can clearly be seen.

\section{PIP vs inverse completeness}\label{sec PIPvscomp}

One interesting question to be answered is wether it possible to take advantage of the completeness map $c(\text{RA},\text{Dec})$  to reduce the variance of the PIP weights.
The idea is that clustering-independent fluctuations such as those purely coming from gaps and tiling overlaps can be safely removed from the budget by defining an effective volume, which, as such, can be traced by the random sample.  
Since the random sample is in general much denser than the galaxy catalogue, weighting randoms has significantly lower effect on the total noise than weighting galaxies.  
Specifically, we can think of creating a random sample with angular density that mimics that of the completeness map while modifying the data pair counts as $\sum w_{ij} \leadsto \sum w_{ij}/(w^{(IC)}_i w^{(IC)}_j)$, where $w_{ij}$ are PIP weights and $w_i^{(IC)}=1/c_i$ the inverse of the completeness map at the position of the $i^{th}$ galaxy.
This procedure does not affect the mean but, if the completeness fluctuations are relevant, has the potential of reducing the overall variance in the correlation function. 

\begin{figure*}
 \begin{center}
   \includegraphics[width=7.0cm]{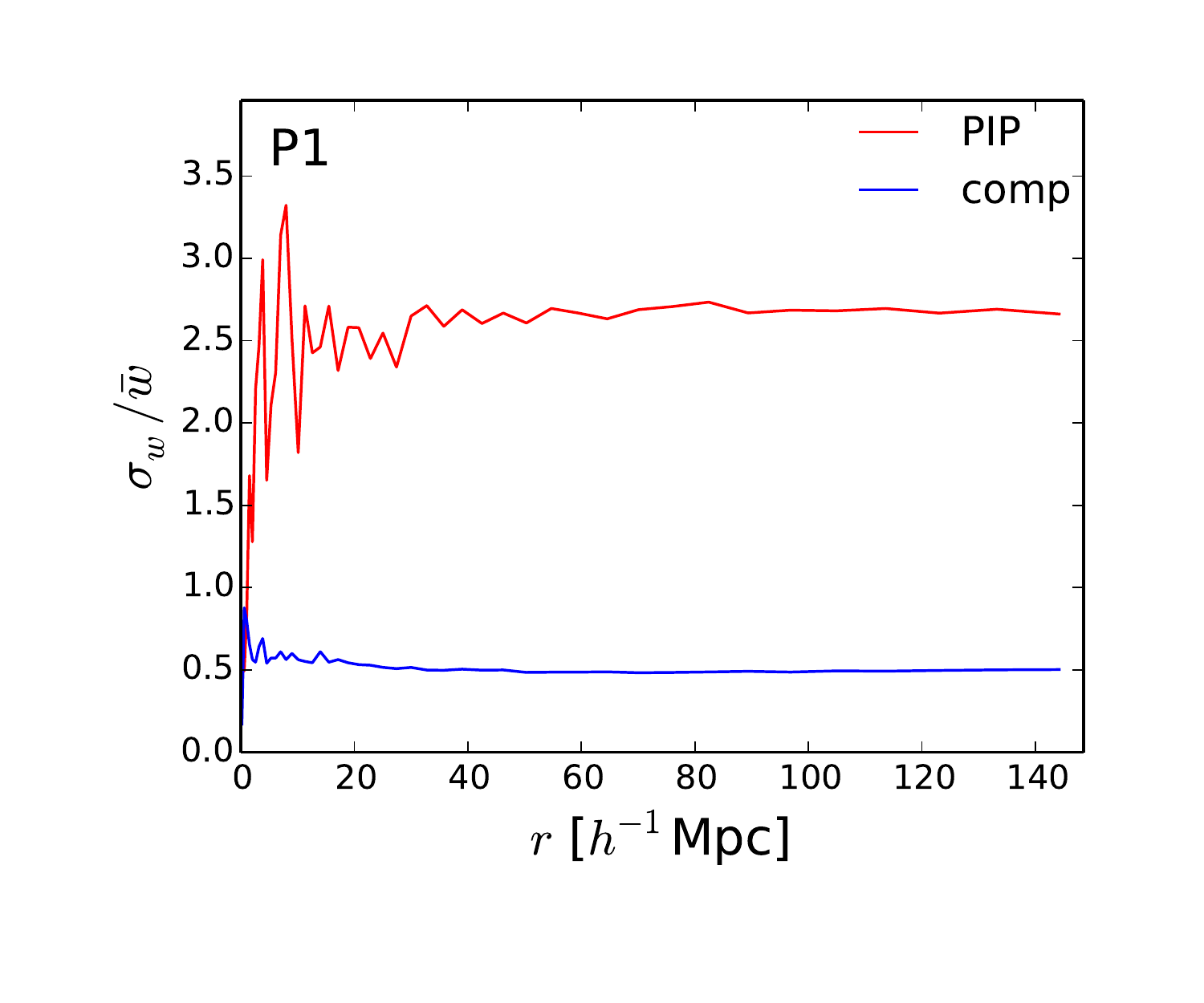}
   \includegraphics[width=7.0cm]{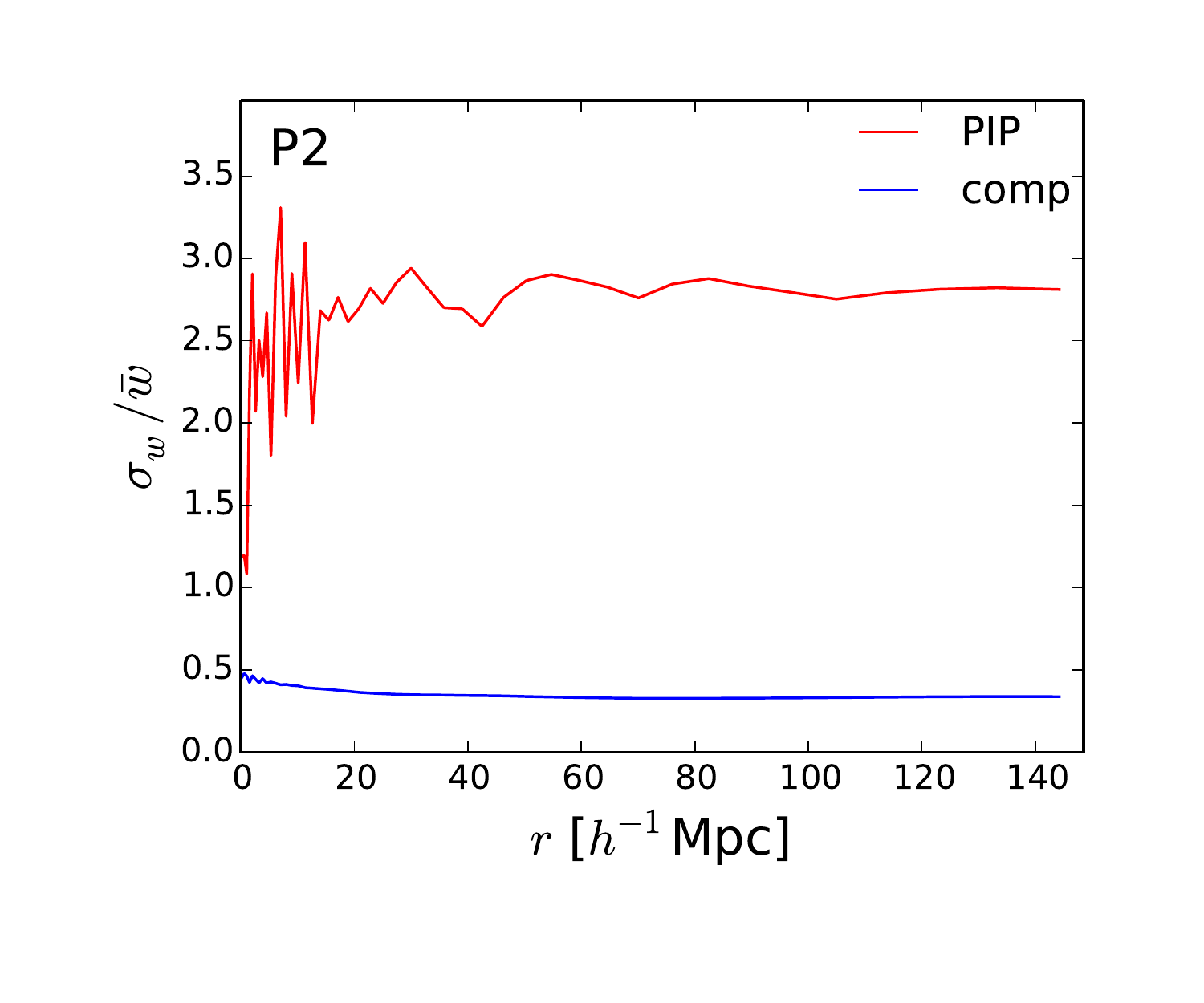}\\
   \includegraphics[width=7.0cm]{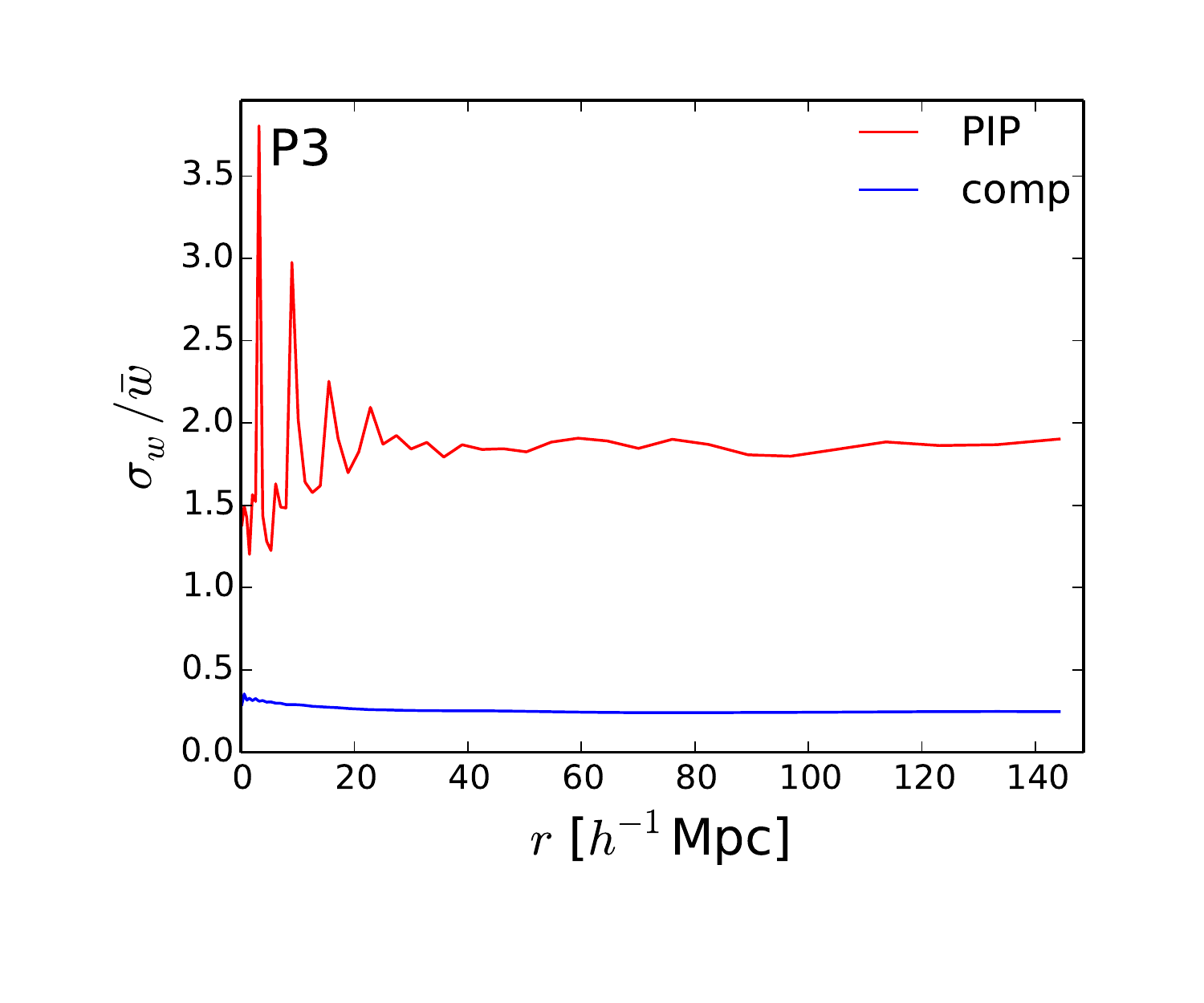}
   \includegraphics[width=7.0cm]{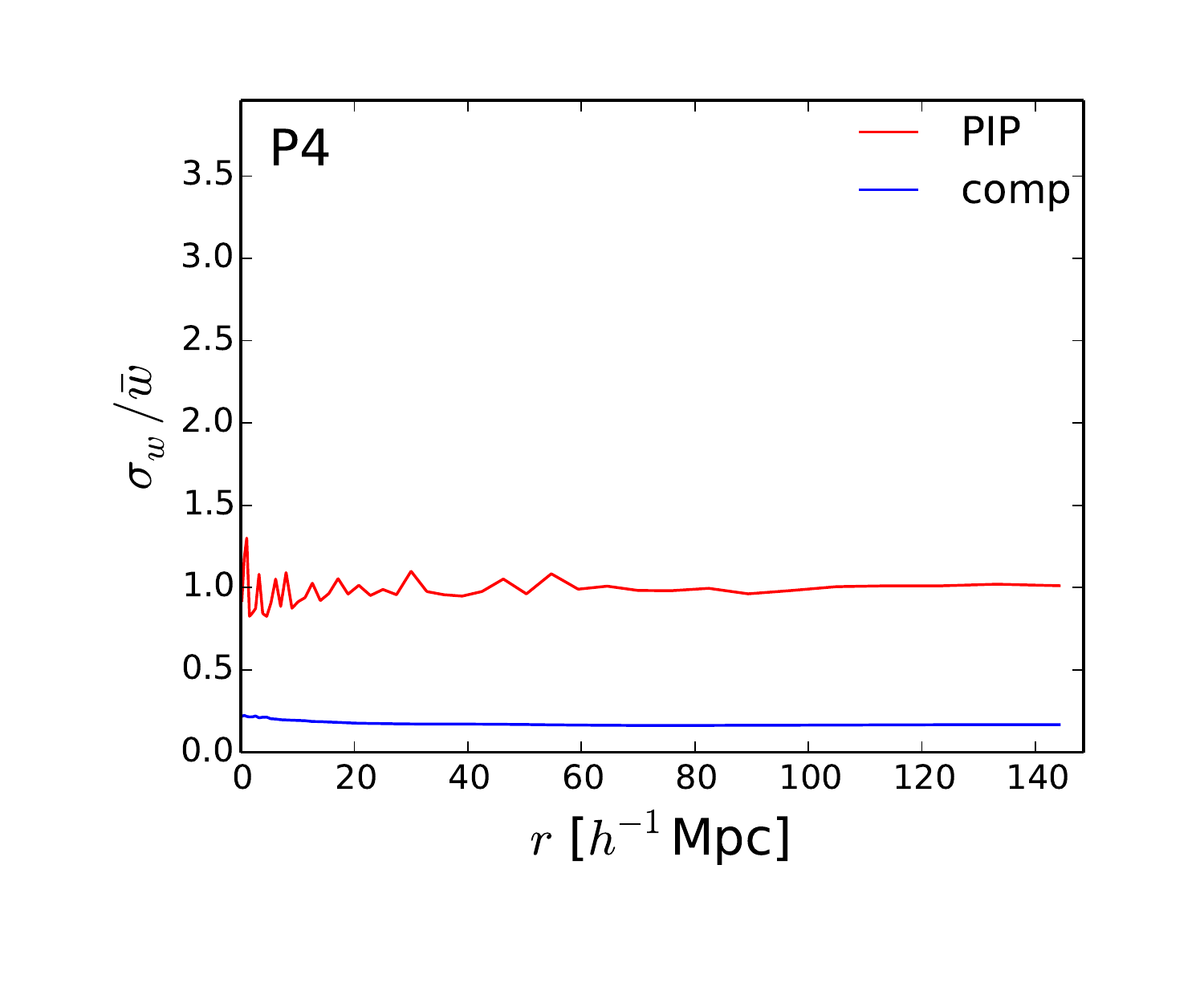}
   \caption{Standard deviation (divided by the mean) of the PIP weights compared to that of the inverse completeness weights, as a function of scale.}
   \label{fig PIPvscomp}
 \end{center}
\end{figure*}

In Fig. \ref{fig PIPvscomp} we show that the contribution coming from the completeness map is actually subdominant on all scales and at any stage of the survey.
Hence, we do not see any advantage in using this map in combination with the PIP correction.   
Note that, for clarity, the PIP variance reported in the figure does not include angular upweighting, the effective variance of the pair counts get significantly smaller when this latter is taken into account as, e.g., for the measurements reported in Fig. \ref{fig vamo}.  
Nonetheless, angular upweighting by construction does not depend on the completeness and, as a consequence, including it would not change our conclusions. 

\bsp	
\label{lastpage}
\end{document}